\documentclass[10pt]{article}

\usepackage{amsmath,amssymb,amsfonts}
\usepackage{graphicx}

\newcommand{\myfig}[3]{
\begin{figure}
\centering
\includegraphics[width=#2cm]{#1.pdf}\caption{#3}\label{#1}
\end{figure}
}

\begin{document}

\newcommand{\N}{N'}
\newcommand{\bC}{{\mathbb C}}
\newcommand{\bZ}{{\mathbb Z}}
\newcommand{\extd}{{\rm d}}

\def\CH{{\cal H}}
\def\CO{{\cal O}}
\def\tr{\hbox{tr}}
\def \CMP {{Commun. Math. Phys.}}
\def \PRL {{Phys. Rev. Lett.}}
\def \PL {{Phys. Lett.}}
\def \NPBProc {{Nucl. Phys. B (Proc. Suppl.)}}
\def \NP {{Nucl. Phys.}}
\def \RMP {{Rev. Mod. Phys.}}
\def \JGP {{J. Geom. Phys.}}
\def \CQG {{Class. Quant. Grav.}}
\def \MPL {{Mod. Phys. Lett.}}
\def \IJMP {{ Int. J. Mod. Phys.}}
\def \JHEP {{JHEP}}
\def \PR {{Phys. Rev.}}
\def \JMP {{J. Math. Phys.}}
\def \GRG{{Gen. Rel. Grav.}}

\newcommand{\ack}[1]{[{\bf Pfft!: {#1}}]}

\newcommand{\pa}{\partial}
\newcommand{\eref}[1]{(\ref{#1})}
\newcommand{\beq}{\begin{equation}}
\newcommand{\eeq}{\end{equation}}
\newcommand{\beqn}{\begin{eqnarray}}
\newcommand{\eeqn}{\end{eqnarray}}

\rightline{hep-th/0604060} \rightline{ILL-(TH)-06-01}
\rightline{VPI-IPPAP-06-04}

\vskip 0.75 cm
\renewcommand{\thefootnote}{\fnsymbol{footnote}}
\centerline{\Large \bf On the Glueball Spectrum of Pure Yang-Mills Theory in $2+1$ Dimensions }
\vskip 0.75 cm

\centerline{{\bf Robert G. Leigh${}^{1}$\footnote{rgleigh@uiuc.edu},
Djordje Minic${}^{2}$\footnote{dminic@vt.edu}
and Alexandr Yelnikov${}^{2}$\footnote{yelnykov@vt.edu}
}}
\vskip .5cm
\centerline{${}^1$\it Department of Physics,
University of Illinois at Urbana-Champaign}
\centerline{\it 1110 West Green Street, Urbana, IL 61801-3080, USA}
\vskip .5cm
\centerline{${}^2$\it Institute for Particle Physics and Astrophysics,}
\centerline{\it Department of Physics, Virginia Tech}
\centerline{\it Blacksburg, VA 24061, U.S.A.}
\vskip .5cm

\setcounter{footnote}{0}
\renewcommand{\thefootnote}{\arabic{footnote}}

\begin{abstract}
We present details of the 
analytic computation of the spectrum of lowest spin glueballs in pure Yang-Mills theory in 2+1 dimensions. The new
ingredient is provided by the conjectured new non-trivial expression for the (quasi)Gaussian part of
the ground state wave-functional. We show that this wave-functional can be derived by solving the Schr\"odinger equation under certain assumptions. The mass spectrum of the theory is
determined by the zeros of Bessel functions, and the agreement with available lattice data is excellent.
\end{abstract}

\newpage

\section{Introduction and Summary}

The understanding of the non-perturbative dynamics of pure Yang-Mills theory
is one of the outstanding problems of theoretical physics.
As was realised a long time ago, one of the persistent difficulties hampering any progress is the fact that we are studying a gauge theory, and it is clear that the fields that we use to define the
theory (in the UV) do not create physical states, and are thus not the ``correct degrees of freedom''. Instead one should switch to some set
of gauge invariant variables. Traditionally, this has meant the Wilson loop operators, and the consequent introduction of the loop space
formalism \cite{polyakov, collective, polyakov1}. Although in some heuristic sense, this would seem to bear some relation to the expected
appearance of a string theory, it is extremely difficult to proceed beyond a few basic stages. We believe that although the Wilson loop plays
a central role in the theory, as its expectation value is a useful order parameter for confinement, it should not be thought of as representing
the ``correct degrees of freedom''.

Instead, building on a construction\footnote{At the moment Karabali-Kim-Nair construction applies only to $2+1$ dimensional Yang-Mills theory. There is
however a close
relation between the Karabali-Kim-Nair formalism and that of Bars
\cite{bars}. This will be explored in Ref. \cite{withlaurent} in
the context of pure Yang-Mills theory in $3+1$ dimensions.} of
Karabali, Kim and Nair \cite{knair}, we suggest that the theory
may be discussed in terms of local gauge invariant variables. The idea of reformulating Yang-Mills theory in terms of local gauge invariant
variables is certainly not new and many such proposals exist in the literature \cite{others, bars}. The unique advantage of Karabali-Kim-Nair formulation however is that
it is possible to extend actual computations to the point when definite {\it quantitative} predictions can be made. For example, in \cite{knair}
string tension in pure YM theory in $2+1$ dimensions has been computed which argees beautifully with numerous lattice simulations \cite{teper, bringoltz}.

Another central element of \cite{knair} is the use of a Schr\"odinger/Hamiltonian approach. Such a formalism \cite{wf,kogan} is not often used in quantum field theory because it is fraught
with regularization issues, and it is somewhat of an art to navigate through them. We believe however that Hamiltonian formalism together with
gauge unvariant local variables make a powerfull combination  which may help to 
uncover many interesting and non-trivial aspects of nonperturbative dynamics of Yang-Mills theories.

In the present paper we wish to make the next step in the Karabali-Kim-Nair program and address the question of determination of glueball mass spectrum in
pure $2+1$ dimensional Yang-Mills theory\footnote{A brief summary of our main results can be found in our recent publication \cite{shortpaper}.}. 
In principle, masses of glueball states can be extracted from the exponential fall-off of vacuum correlators of various gauge invariant
probe operators and this is the point of view we take in this paper. This technique however requires sufficient knowlege of the vacuum of the
theory. We believe that it is useful to approximate the vacuum wave functional as a generalized\footnote{This notion will be made precise below.}
quasi-Gaussian in a certain variable (appropriate to a ``correct degree
of freedom'' describing fluctuations around the vacuum) but with a
very non-trivial kernel which contains information about an
infinite number of physical states. It is in this sense that a
simple idea like a quasi-Gaussian wave functional, reminiscent of a
``constituent picture'' of glueballs \cite{constituent}, is
capable of encoding a stringy spectrum. We note in passing that
many of these ideas have analogues in condensed matter physics
(e.g., superfluids, superconductors and quantum Hall fluids
\cite{fradkin}). The resulting picture of glueballs is
more reminiscent of open strings, as opposed to the closed-string
picture suggested by Wilson loops.

We want to be absolutely clear that the
main outcome of our analysis in this paper is a proposal for a
new non-trivial expression for the vacuum wavefunctional.
As we show in Chapter~3 this wavefunctional can be derived by solving the functional
Schrodinger equation to quadratic order in local gauge invariant variable $\bar\partial J$ \footnote{This variable can roughly be thought of as magnetic field $B$.
See equation (\ref{BJ}) below.} of Karabali, Kim and Nair
{\it and} under a certain extra assumption on the spectrum of the kinetic energy operator. At the moment this latter assumption should
be considered as {\it conjectural} however and, at least in principle, it should be possible to prove (or disprove) it by direct
computation\footnote{Preliminary analysis of the issues involved is presented in Appendix A.}.

It would be difficult to judge the validity of our construction were
it not for the existence of ``experimental'' lattice data. Pure Yang-Mills theory on the lattice has been studied primarily by Teper
and collaborators \cite{teper}, and information on a significant number of low lying states with a variety of Lorentz quantum numbers is available.
Although we have attempted in this paper to provide as much detail as possible, there remain a number of outstanding conjectural results that require
further research, but it should be clear to the reader that our analytic results agree extremely well with the lattice data. This agreement is non-trivial
on many accounts, as we will explain below.

One topic that we will not touch on in this paper is the role of topology.  Clarifying this point would certainly be of interest. In particular,
a variety of (non-gauge-invariant) configurations have, over the years, been proposed as relevant to confinement. We should note though
that there is a crucial entry for topology in confining theories, that of the ``compactness'' of the configuration space leading to a discrete
spectrum \cite{twoplusone, greensite, gribov}. The interpretation of the kinetic energy operator as a Laplacian on a compact configuration space
has been discussed previously in \cite{twoplusone, knair}, and is an essential motivational part of our analysis in this paper.

Finally, in principle our vacuum wave functional is good (as a
first approximation to the lattice data) for any rank $N$ of the gauge group. Thus the ratios of
glueball masses to the string tension (string tension being different
for different ranks of the gauge group) approximate the
available lattice data reasonably well for any $N$. However, as was found
previously by Karabali-Kim-Nair, and is confirmed by our analysis, the string tension compares best with
lattice simulations at large $N$, and because of this observation, we
expect that predictions for the glueball masses, as implied by
the vacuum wave functional discussed in this paper, work best at large $N$ as well.

We have organized the paper as follows. In
Section~\ref{sec:formalism}, we provide a review of the
Karabali-Nair parameterization of $2+1$ dimensional Yang-Mills
theory, emphasizing the aspects that are most important for our
construction. In particular, we focus on the appearance of
holomorphic symmetry, which plays a central role in the theory. In
a subsection, we also provide a discussion of the action of spin,
parity and charge conjugation in these variables, which is crucial
for both the construction of the vacuum wave functional as well as
the correct identification of physical states. We also give some
details of the Hamiltonian in these variables, and how the theory
can be regulated in a gauge and holomorphic invariant fashion. In
Section~\ref{sec:vacwf}, we discuss our proposal for the vacuum wave
functional, its likely validity and some of its physical
consequences. We then discuss the Schr\"odinger equation and its
solution. As was mentioned above this part of the analysis contains many regularization-related subtleties which have not yet been completely clarified.
The form of the Schrodinger equation that we employ should thus be considered preliminary; however, the central point of the present paper is simply that the solution of this Schrodinger equation fits lattice data remarkably well. Further clarification of these results, we hope, will be uncovered in future work. Four Appendices are also provided as supplementary
material for this discussion. In Appendix~A, we give some detail on the computations required for the Schr\"odinger equation, focussing on the difficult regulator issues. In Appendix~B, we
compute the (divergent) vacuum energy, and in Appendix~C, we
discuss the general solution of the Schr\"odinger equation. It is
in the later discussion that we see a crucial result: {\it the
only normalizable solution to the Schr\"odinger equation
corresponds to the confining  vacuum.} In fact, the vacuum wave
functional can be thought of as interpolating between perturbative regime
 in the UV and the confining physics of the
IR. 
Appendix~D
clarifies the appearance of glueballs as single particle
asymptotic states. In Section~4, we discuss invariant probe
operators of definite $J^{PC}$ quantum numbers whose correlation
functions can be used to extract the mass spectrum of glueball
states. Because of the non-trivial quasi-periodic structure of the
kernel appearing in the vacuum wave-functional, we find an
infinite number of such excitations. In this section we also
investigate more fully the spectrum of low-lying glueball states,
their approximate degeneracies and the corresponding Regge
trajectories. We conclude this paper with a series of open
questions that we have left for future work.

\section{Formalism}\label{sec:formalism}
Consider pure $SU(N)$ Yang-Mills theory in $2+1$ dimensions in the Hamiltonian gauge $A_0=0$. It is convenient to parameterize the spatial coordinates in terms
of complex variables $z=x_1-i x_2$ and $\bar{z} =x_1+i x_2$, and we will write $\partial \equiv \partial/\partial z$, $\bar\partial \equiv \partial/\partial\bar z$. The spatial components of the gauge field may be written
\begin{equation}
A = \frac12 (A_1+iA_2),\ \ \ \ \ \bar{A} = \frac12 (A_1-iA_2).
\end{equation}
We will frequently use the notation $A_{1,2}=-it^aA_{1,2}^a$ where $t^a$ are $N\times N$ matrices representing the $SU(N)$ Lie algebra $[t^a, t^b] = if^{abc} t^c$ with
normalization ${\rm Tr}(t^at^b) =\frac12\delta^{ab}$. When appropriate, we will use a specific complex index notation for the components of these matrices, $t^a_{i\bar j}$, with $i,j=1,...,N$.

The quantization of this theory can be considered within the Hamiltonian formalism. Our approach will be based upon a change of variables, whose many details have been worked out by Karabali, Kim and Nair \cite{knair}. The Karabali-Nair parameterization is
\begin{equation} \label{hl}
A_{i\bar j}= - \sum_\alpha\partial M_{i\bar\alpha} (M^{-1})_{\alpha\bar j}\,, \quad \bar{A}_{i\bar j}= + \sum_\alpha (M^{\dagger -1})_{i\bar\alpha}\bar\partial
(M^{\dagger})_{\alpha\bar j}\,.
\end{equation}
Here, $M$ is an invertible complex matrix variable, whose index structure we have denoted explicitly; generally, we will simplify notation by dropping
explicit indices and ordering the expressions appropriately. The tracelessness of the gauge field corresponds to the unimodularity of $M$ and so $M\in SL(N,\bC)$.
Note that a (time independent) gauge transformation
\begin{equation}
A\ \mapsto\ g A g^{-1} - \partial g g^{-1},\ \ \ \
\bar A\ \mapsto\ g \bar A g^{-1} - \bar\partial g g^{-1},
\end{equation}
 where $g\in SU(N)$ becomes simply $M \mapsto g M$.
Correspondingly, a {\it local gauge invariant} variable is
\begin{equation}
H \equiv  M^{\dagger} M.
\end{equation}

The definition of $M$ implies a {\it holomorphic invariance}
\begin{equation}\label{eq:holotrans}
\begin{array}{rcl}
M(z,\bar z) &\mapsto& M(z,\bar z)h^\dagger (\bar{z})\vspace{0.1in}\\
M^\dagger(z,\bar z) &\mapsto& h({z})M^\dagger(z,\bar z)
\end{array}
\end{equation}
where $h(z)$ is an arbitrary unimodular complex matrix whose matrix
elements are independent of $\bar z$. This is distinct from the original
gauge transformation, since it acts as right multiplication rather than
left and is holomorphic. Under the holomorphic transformation, the gauge
invariant variable $H$ transforms homogeneously
\begin{equation}
\label{holo}
H(z,\bar z)\ \mapsto\ h(z) H(z,\bar z) h^\dagger(\bar{z}).
\end{equation}
The theory written in terms of the gauge invariant $H$ fields will have
its own local (holomorphic) invariance. In a sense, we have replaced one local invariance by another.
The gauge fields, and the Wilson
loop variables, know nothing about this extra invariance. We will deal
with this by requiring that the physical state wave functionals be
holomorphically invariant.

One of the most extraordinary properties of this parameterization however is
that the Jacobian relating the measures on the space of connections ${\cal C}$
and on the space of gauge invariant variables $H$ can be explicitly
computed \cite{knair}
\begin{equation}
d \mu [{\cal C}] = \sigma\, d \mu [H]\, e^{2c_A S_{WZW}[H]}
\end{equation}
where $c_A$ is the quadratic Casimir in the adjoint representation of
$SU(N)$ ($c_A = N$), $\sigma$ is a constant determinant factor and
\begin{equation}
S_{WZW}[H] = \frac{1}{2\pi}\int\!\! d^2z\
{\rm Tr} \left(\partial H \bar\partial H^{-1}\right)
\,+\frac{i}{12\pi} \int\!\! d^3x\ \epsilon^{\mu\nu\lambda}
{\rm Tr} (H^{-1}\partial_\mu HH^{-1}\partial_\nu HH^{-1}\partial_\lambda H)
\end{equation}
is the  Wess-Zumino-Witten (WZW) action for the hermitian matrix field $H$, which is both
gauge and holomorphic invariant. The quantum inner product may be written
as an overlap integral of gauge invariant wave functionals with
non-trivial measure
\begin{equation}
\langle1|2\rangle= \int d \mu [H]\: e^{2c_A S_{WZW}(H)}\: \Psi^*_1 \Psi_2.
\end{equation}
From many of the above expressions, it is clear that a useful gauge-invariant
variable is the ``current''
\begin{equation}
J
=\frac{c_A}{\pi}\,
\partial H H^{-1}.
\end{equation}
In particular, it is easily established that the standard Wilson loop operator in the fundamental representation of $SU(N)$ may be written in terms of $J$ as
\begin{equation}\label{eq:WilsonLoop}
\Phi(C)={\rm Tr}\: P\, {\rm exp} \left(\oint_C dz\ \partial H H^{-1}\right) = {\rm Tr}\: P\, {\rm exp} \left(\frac{\pi}{c_A}\oint_C dz\ J\right)  .
\end{equation}
Under the holomorphic transformations  $J$
transforms as a holomorphic connection
\begin{equation}\label{eq:currtrans}
J\ \mapsto\  hJh^{-1}+\frac{c_A}{\pi}\,
\partial h h^{-1}
\end{equation}
while $\bar\partial J$ transforms tensorially, $\bar\partial J\mapsto h(z)\bar\partial J h^{-1}(z)$. Consequently, $\bar\partial J$ plays a central role in this theory; indeed, it is closely related to the magnetic field
\begin{equation}\label{BJ}
B = -\frac{2\pi}{\, c_A}\,
M^{\dagger -1} \bar{\partial}J M^{\dagger} .
\end{equation}
As is clear from (\ref{eq:currtrans}), the current is a connection for holomorphic transformations and there is a corresponding covariant derivative. For example, for an adjoint tensor field $\phi_{\alpha\bar\beta}$, we have
\begin{equation}
[D,\phi]=\partial\phi-\frac{\pi}{c_A}\,
[J,\phi]
\end{equation}
which also transforms homogeneously under (\ref{eq:holotrans}).
Furthermore, note that we may define an antiholomorphic current
\begin{equation}
\bar J=\frac{c_A}{\pi}\,
\bar\pa HH^{-1}
\end{equation}
which is distinct from the adjoint of $J$; in fact $J^\dagger=H^{-1}\bar JH$. It is easily shown that
\begin{equation}
[D,\bar J]=\bar\partial J
\end{equation}
which may be regarded as a reality condition on $J$.
Using this, we see that
\begin{equation}
(\bar\pa J)^\dagger=\pa J^\dagger=\pa (H^{-1}\bar JH)=H^{-1}[D,\bar J]H=H^{-1}(\bar\pa J)H.
\end{equation}
Thus apart from conjugation by $H$, $\bar\pa J$ is essentially real. This is one of the reasons why it is possible to write the theory entirely in terms of $J$. For notational purposes, we introduce the covariant Laplacian
\begin{equation}\label{eq:covLaplace}
\Delta=\frac{\bar\pa D+D\bar\pa}{2}.
\end{equation}
We note that $\bar\pa D-D\bar\pa=-(\pi/c_A)\,\bar\pa J$.

We will also need a variety of fields written in the adjoint representation. Suitable formulas are such as
\begin{eqnarray}
M^{ab}\!&=&\! 2\,{\rm Tr}( t^a M t^b M^{-1})\\\nonumber
H^{ab}\!&=&\! 2\,{\rm Tr}( t^a H t^b H^{-1}).
\end{eqnarray}
One may check that these expressions do indeed transform under holomorphic transformations in the adjoint representation.

To conclude this section, we summarize by noting that we may parameterize everything in this theory, including the path integral measure,  in terms of gauge invariant variables, but in doing so we encounter a new holomorphic invariance which is not seen by the original gauge fields. It is fundamental to this theory that we take this new invariance into account.

\subsection{Spacetime Quantum Numbers}

We will often be interested in gauge and holomorphic invariant operators constructed as traces of products of $\bar\partial J$ and its derivatives. As it will be necessary to classify such operators with respect to spacetime quantum numbers $J^{PC}$, we pause to discuss these now.

Here $J$ is the quantum number associated with the spatial $SO(2)$ subgroup of the Lorentz group. Clearly, $A$ ($\bar A$) carries $SO(2)$ charge $-1$ ($+1$), and derivatives carry spin $J_\partial=-1$, $J_{\bar\partial}=+1$. Upon the KN reparameterization, $H$ will have spin zero, and thus the current $J$ carries angular momentum $J_J=-1$. Consequently, $J_{\bar\partial J}=0$. For the most part then, the spin of an invariant operator will be determined by the net number of derivatives.

Parity and charge conjugation are determined as follows. By parity, we will mean the operation
\begin{equation}
x_1\: \stackrel{P}\mapsto\: x_1,\ \ \ \ \ x_2\: \stackrel{P}\mapsto\: -x_2.
\end{equation}
Thus, under $P$ we have
\begin{equation}
\begin{array}{llrcl}
P:&&z\!\! &\mapsto&\!\! \bar z\\
&&A\!\! &\mapsto&\!\!  \bar A\\
&&M\!\! &\mapsto&\!\! M^{\dagger -1}\\
&&H\!\! &\mapsto&\!\!  H^{-1}\\
&&\bar\pa J\!\! &\mapsto&\!\! -H^{-1}\bar\pa J H\\
&& \Delta\!\! &\mapsto&\!\! H^{-1}\Delta H.
\end{array}
\end{equation}
A field which is conjugated by $H$ (up to sign) under parity is transforming tensorially
\begin{equation}
\Phi\: \stackrel{P}\mapsto\: \alpha_{\Phi} H^{-1}\Phi H
\end{equation}
where $\alpha_{\Phi} = \pm 1$. We then find
\begin{equation}
 [D,[\bar\pa, \Phi]]\ \stackrel{P}\mapsto\ \alpha_{\Phi} H^{-1}[\bar\partial, [D,\Phi]] H
\end{equation}
and
\begin{equation}
[\bar\partial, [D,\Phi]]\ \stackrel{P}\mapsto\ \alpha_{\Phi} H^{-1}[D,[\bar\pa, \Phi]] H.
\end{equation}
Derivative operators of definite parity will thus be even(odd) linear combinations of these two. Taking the sum of the two, we construct
\begin{equation}
[\Delta,\Phi]\equiv \frac{1}{2}\left( [D,[\bar\pa,\Phi]]+[\bar\pa,[D,\Phi]]\right)
\end{equation}
and
\begin{equation}
P:\ \ [\Delta,\Phi]\ \mapsto\ \alpha_\Phi [\Delta,\Phi].
\end{equation}
Note in particular that $\Delta^{\!n}\Phi$ has the same parity as $\Phi$.

Charge conjugation does not act spatially ($z\to z$), but we must have
\begin{equation}
C:\quad
A_{i\bar j}\ \mapsto\  -A_{j\bar i}\qquad
\bar A_{i\bar j}\ \mapsto\  -\bar A_{j\bar i}.
\end{equation}
For $M$ and $M^\dagger$, a choice of action of charge conjugation consistent with this is
\begin{equation}
M_{i\bar\alpha}\ \stackrel{C}\mapsto\ (M^{-1})_{\alpha\bar i}\qquad
M^\dagger_{\alpha\bar i}\ \stackrel{C}\mapsto\ (M^{\dagger -1})_{i\bar\alpha}
\end{equation}
which leads to
\begin{equation}
H_{\alpha\bar\beta}\ \stackrel{C}\mapsto\  (H^{-1})_{\beta\bar\alpha}\qquad
J_{\alpha\bar\beta}\ \stackrel{C}\mapsto\ -J_{\beta\bar\alpha}.
\end{equation}
Note that generally if
\begin{equation}
C:\ \ \phi_{\alpha\bar\beta}\ \mapsto\ \phi^C_{\beta\bar\alpha}
\end{equation}
 then
\begin{equation}
C:\ \ ([D,\phi])_{\alpha\bar\beta}\ \mapsto\ +([D,\phi^C])_{\beta\bar\alpha}.
\end{equation}
 Thus $C$ counts $J$'s (mod 2) that are not inside $D$'s.

\subsection{The KN Hamiltonian}

The standard $YM_{2+1}$ Hamiltonian
\begin{equation}
{\cal H}_{YM}\equiv T + V = \int {\rm Tr} \left(g_{YM}^2 {E_i}^2 + \frac{1}{g_{YM}^2} {B}^2\right)
\end{equation}
where as usual, the electric fields $E_i$ play the role of momenta
conjugate to $A_i$. The Hamiltonian can be rewritten explicitly in
terms of gauge invariant variables. This was worked out in detail
by Karabali and Nair, and has the {\it collective field form}
\begin{eqnarray}\label{Hamilt}
{\cal H}_{KN}[J]=
m \left(\int_x J^a(x) \frac{\delta}{\delta J^a(x)} +
\int_{z,w}\Omega^{ab}(z,w)
\frac{\delta}{\delta J^a(z)} \frac{\delta}{\delta J^b(w)}\right) +
\frac{\pi }{mc_A} \int_x \bar{\partial} J^a \bar{\partial} J^ a
\end{eqnarray}
where
\begin{equation}
m = \frac{g_{YM}^2 c_A }{2 \pi}\end{equation}
and in most cases $\Omega^{ab}$ can be thought of as
\begin{equation}\label{eq:Omegasimple}
\Omega^{ab}(z,w) = \frac{c_A}{\pi} D_w^{ba} \bar G(w-z).
\end{equation}
Here $\bar G(w-z)$ is the ordinary Green's function defined by $\bar\partial_w \bar G(w-z) = \delta^{(2)}(w-z)$ and
\begin{equation}
D_w^{ba} = \delta^{ba}\partial_w - i f^{bda}\frac{\pi}{c_A}\, J^d(w).
\end{equation}
In what follows, however, we will consider the action of
${\cal H}_{KN}$ on local gauge invariant operators, and for such purposes a more general point-split expression for $\Omega^{ab}$
will be needed (see \cite{knair} or Appendix~A for details).

The derivation of this Hamiltonian involves carefully regulating certain
divergent expressions in a gauge invariant manner \cite{knair}. We note that the scale $m$ is essentially the 't~Hooft
coupling \cite{largeN}. 
\section{Vacuum Wave Functional}\label{sec:vacwf}

Our basic goal is to determine the masses of some of the lowest
lying glueball states.  The technique that we will use is to consider vacuum
correlation functions of invariant operators at large spatial separation. Consequently we wish to determine the form of the
vacuum wave functional. Of course, the determination of the full expression for the ground state is an insurmountable problem.
Therefore our goal here is more modest: we wish to determine only the quasi-Gaussian part of the vacuum wave functional. Two comments
about this program are in order.

First, the main outcome of our analysis\footnote{The main outcome of our analysis is summarized in eqs. (\ref{eq:psians}) and (\ref{eq:normvacuum}).} in this chapter should be considered as conjectural. The derivation of the
quasi-Gaussian part of the vacuum wave-functional that we are going to present below depends in a crucial way on a certain
assumption\footnote{See eq. (\ref{eq:TOn}) and further comments that follow it.} about the  spectrum of the kinetic energy operator. At the moment we {\it cannot}
prove that this assumption is correct,
and therefore all results that follow from it should be considered as preliminary.

Second, the next natural question to ask here is to what extent (or in which regime) this  quasi-Gaussian expression approximates (if it is an approximation at
all!) the true vacuum of the
theory. Unfortunately, at the moment this question can be answered only {\it a posteriori}. In particular, as we will see in the next section, estimates of 
glueball  masses based on our wavefunction reproduce the lattice data remarkably well. 

In this respect we want to point also that a useful insight on the possible range of applicability of our wave functional comes if we take $1+1$ dimensional Yang-Mills theory coupled to
adjoint matter\footnote{This theory is related to pure glue case in $2+1$ dimensions by dimensional reduction on a circle.} as a guide.
A version of this theory has been studied in the light-cone formulation \cite{Demeterfi:1993rs};
in that context, partons play the role of constituents, and numerical work on the lowest-lying glueball states showed that
the wave functions of the low lying "glueball" states have probability very close to one of being an eigenstate of parton number operator.
The relation between the partons of this theory and the constituent glue in our study is not clear, nevertheless if we take the results of 
\cite{Demeterfi:1993rs} as a guide, 
we then expect that the constituent picture
which emerges within our scheme (and which is similar to parton picture of \cite{Demeterfi:1993rs}) should give a reliable description of low lying glueball states.

After these preliminary discussion we are ready to proceed to details of our proposal. If we look at the form of the KN Hamiltonian, we note that if we were to drop the potential term (for example,
this would happen in the large $m$ limit), then $\Psi=1$ would satisfy the Schr\"odinger equation. Furthermore,
we note that, given the non-trivial form of the measure, this wave functional would be normalizable! Consequently,
we are motivated to consider a ground state wave functional of the form $\Psi_0=e^P$, and we would like to
determine $P$. In principle, $P$ can be any functional which is gauge and holomorphic invariant, as well as invariant
under spacetime symmetries.

Of course, finding an exact expression for $P$ is a very difficult task. However, as we said above, we want to find $P$ to quadratic 
order only. Therefore one might consider a wave functional that is Gaussian in $\bar\pa J$
\begin{equation}\label{eq:psiGaussian}
\Psi_0 = \exp\left( -
\frac{\pi}{2 c_A m^2}
\int \bar{\partial} J^a\ K\left(\frac{\!\partial\bar\partial}{\,m^2}\right) \bar{\partial} J^a +\ldots\right)
\end{equation}
containing a kernel $K$ that we wish to determine.

The form of an ansatz (\ref{eq:psiGaussian}) is reminiscent
of what one usually takes in variational calculations. We want to
be clear, however, that our approach is not variational in the
sense that we are not going to minimize vacuum energy density with
this ansatz. Our intension is to find the form of the kernel
$K(\partial\bar\partial /m^2)$ by trying to actually solve the Schr\"odinger
equation to quadratic order in $\bar\pa J$. In other words we are attempting to
really find an explicit form of $P$ to that order in $\bar\pa J$.

To do this properly, however, requires a further generalization as (\ref{eq:psiGaussian}) is
not invariant under holomorphic transformations
(\ref{eq:holotrans}). 
To repair this problem, we
take the generalized quasi-Gaussian
\begin{equation}\label{eq:psians}
\Psi_0 = \exp\left( -
\frac{\pi}{2 c_A m^2}
\int \bar{\partial} J\ K\left(\frac{\Delta}{m^2}\right) \bar{\partial} J +\ldots\right)
\end{equation}
which is explicitly gauge and holomorphic invariant. Here $\Delta=\{D,\bar\pa\}/2$ is the holomorphic covariant Laplacian (\ref{eq:covLaplace}) mentioned earlier, which depends on $J$. Note also that the argument of the exponential is real and invariant under
spin, parity and charge conjugation. It contains an as yet arbitrary dependence on $\Delta$ but is quadratic in the commutator $\bar\pa J$. From the results of section 2 on discrete symmetries, it makes sense to organize the wave-functional in this way. The ellipsis in  (\ref{eq:psians}) would then contain terms of quartic order and higher in  $\bar\pa J$. As we will see, the Schrodinger equation can also be organized along these lines, and it will be sufficient for our purposes to consider the terms explicitly shown in (\ref{eq:psians}).

It is convenient to write $K(L)$, with $L=\Delta/m^2$, as a formal infinite power series
\begin{equation}\label{KTaylor}
K(L) = \sum_{n=0}^{\infty} c_n L^n
\end{equation}
with as yet unknown coefficients $c_n$.  We see that the generalized ansatz (\ref{eq:psians}) corresponds to the expansion
\begin{equation}\label{eq:Pansatz}
P=-\frac{\pi}{2 c_A m^2}\sum_n c_n \frac{{\cal O}_n}{m^{2n}} +\ldots
\end{equation}
of $P$ in terms of  gauge and holomorphic invariant operators
\begin{equation}
\CO_n \equiv \int\bar{\partial} J (\Delta^{n})\, \bar{\partial} J.
\end{equation}
Note that of all possible local operators that might appear in $P$ (and which are formally represented by "$\ldots$" in (\ref{eq:psians})
and (\ref{eq:Pansatz})) we keep only this subset
since only these operators contain part quadratic in $\bar\pa J$. In other words, even though the ansatz (\ref{eq:psians}) is not Gaussian
any longer and contains ({\it via} $\Delta$) terms of higher order in $J$, it certainly {\it does not contain all} such terms.
As was mentioned above, we intend to solve the Schr\"odinger equation to quadratic order in $\bar\pa J$ only and thus we keep only those
terms which are required for consistency.

It should be noted that we are taking here a basis of {\it local} operators.\footnote{Nevertheless, the form of $K$ that we will arrive at later is an infinite power series and is thus non-local.} As we have seen, this basis is natural from the point of view of holomorphic and discrete symmetries. It is far from clear whether or not this is the most convenient choice, and in particular, we would not expect that this choice diagonalizes the Hamiltonian.  We will discuss many of these and related issues, particularly in Appendix A.

One way to intuitively motivate the expansion in (\ref{eq:psians}) is to think of $\bar\pa J$ as the relevant local
probes of real physical states. Then there should exist an expansion parameter, which is related to the size of the glueballs. The quadratic
term in (\ref{eq:psians}) should be then interpreted as the leading term in the expansion in the inverse of that effective glueball size.
This would be very reminiscent of the $\alpha'$ expansion in string theory.

After such preliminary discussion of the wave-functional and before
proceeding to the Schr\"odinger equation and its solution, let us
say a few words about the (expected) asymptotic behavior of the vacuum
state. In the UV, the vacuum wave functional should correspond to
free gluons \beq \Psi_0^{UV}\ \mapsto\ {\rm exp}\left({-\frac{1}{2
g_{YM}^2}\int  B^a\frac{1}{|p|} B^a}\right). \eeq Because of the
relation (\ref{BJ}) between $B$ and $\bar\pa J$, we see that this
is Gaussian in $\bar\pa J$.

From the form of the vacuum wave functional in the UV, it is clear that one cannot expect $K$ to be a local functional of $L=\Delta/m^2$.
However, we will formally write $K(L)$ as an infinite power series (\ref{KTaylor})
and seek a summation which corresponds to a normalizable solution of the Schr\"odinger equation. As we will see the resulting kernel
$K$ will have a very non-trivial form, which contains much physical information. 

Before proceeding further, let us note that in the small $L$ (IR) limit, we
expect the kernel to asymptote to a constant value. In fact, we will find that
\begin{equation}
\Psi_0^{IR}\ \mapsto\ \exp\left( - \frac{1}{2 g_{YM}^2 m} \int {\rm Tr}\, B^2\right).
\end{equation}
As explained in \cite{knair}, this
wave functional can be thought of as providing a probability measure $\Psi_0^* \Psi_0$
equivalent to the partition function of the Euclidean two-dimensional
Yang-Mills theory with an effective Yang-Mills coupling $g_{2D}^2 \equiv
m g_{YM}^2$. Using the results from \cite{2dym}, Karabali, Kim and Nair
deduced the area law for the expectation value of the Wilson loop
operator (\ref{eq:WilsonLoop})
\begin{equation}
\langle\Phi\rangle \sim \exp( - \sigma A)
\end{equation}
with the string tension following from the results of \cite{2dym}
\begin{equation}\label{stringtension}
\sigma = g_{YM}^4 \frac{N^2-1}{8\pi}.
\end{equation}
This formula agrees nicely  with extensive lattice simulations
\cite{teper}, and is consistent with the appearance of a mass gap
as well as the large $N$ 't~Hooft scaling\footnote{It should be
noted however that this result is not completely satisfactory at finite
$N$: the representation dependence of this result would not be
correct \cite{greensite, chris}; it is however consistent with ``Casimir scaling''. We will not discuss this
important issue (as well as matters relating to the center of the
gauge group) in this paper, but it is one indication that the formulation may only be consistent at large $N$.}.

\subsection{The Schr\"odinger equation}

Now let us return to our discussion of the derivation of the vacuum wave functional. To properly discuss the Schr\"odinger equation, it is necessary to regulate the Hamiltonian, as we have discussed briefly above. It is fairly straightforward to see however, what the general form of the Schr\"odinger equation will be. Given our ansatz, we will find
\begin{equation}\label{eq:SchrEqGen}
{\cal H}_{KN}\Psi_0=E_0\Psi_0=\left[ E_0+\int\bar\pa J\left( {\cal R}\right)\bar\pa J +\ldots\right]\Psi_0 .
\end{equation}
The (divergent) vacuum energy $E_0$ can be isolated, and in Appendix B we give its derivation; as expected, the leading divergence in the UV is cubic. Next, what we need to do is compute the expression that we have labelled by ${\cal R}$ and set it to zero. This will constitute an equation for the kernel $K$. The ellipsis in (\ref{eq:SchrEqGen}) 
stands for the terms which are at least cubic in $\bar\pa J$. Given our discussion above, we can neglect these
terms.

Let us consider the various terms in ${\cal R}$. First, the potential term $\int\bar\pa J\bar\pa J$ clearly contributes a fixed constant to ${\cal R}$.
The remainder of ${\cal R}$ will come from the action of the kinetic energy operator $T_{KN}$. Given the form of the vacuum wave functional $\Psi_0 \sim e^P$, it is elementary
to derive
\beq
\frac{\delta\Psi_0}{\delta J^a(z)}=\frac{\delta P}{\delta J^a(z)}\Psi_0
\eeq
and thus
\beq\label{eq:formalSch}
\frac{\delta^2\Psi_0}{\delta J^a(z)J^b(w)}=\left[\frac{\delta^2 P}{\delta J^a(z)J^b(w)}+\frac{\delta P}{\delta J^a(z)}\frac{\delta P}{\delta J^b(w)}\right]\Psi_0
\eeq
from which we find
\begin{equation}\label{eq:PSchrodinger}
{\cal H}_{KN}\Psi_0 = \left[T_{KN}P+ m\int_{z,w}\Omega^{ab}(z,w) \frac{\delta P}{\delta J^a(z)}\frac{\delta P}{\delta J^b(w)} + \frac{c_A}{\pi m} \int\bar\pa J^a\bar\pa J^a\right]\Psi_0.
\end{equation}
The second term in brackets is easy to compute. Since we want to solve Schr\"odinger equation to quadratic order in $\bar\partial J$ only, it is enough to use ansatz (\ref{eq:psiGaussian}) from which we obtain
\begin{equation}\label{eq:Pquadratic}
P =  - \frac{\pi}{2 c_A m^2}
\int \bar{\partial} J\ K\left(\frac{\partial\bar\partial}{m^2}\right) \bar{\partial} J + \ldots
\end{equation}
and
\begin{equation}
\frac{\delta P}{\delta J^a(z)} = \frac{\pi}{ c_A m^2}\left[ \bar\partial\, K\!\left(\frac{\partial\bar\partial}{m^2}\right)\right]_z \bar{\partial} J^a(z) +\ldots.
\end{equation}
Also to this order we have from (\ref{eq:Omegasimple})
\begin{equation}
\Omega^{ab}(z,w) = \frac{c_A}{\pi} \delta^{ab} \partial_w \bar G(w-z) + \ldots
\end{equation}
and by putting everything together we see  that the second term in brackets in (\ref{eq:PSchrodinger}) is equal to
\begin{equation}
\frac{\pi}{c_A m} \int \bar{\partial} J\left[ \frac{\partial\bar\partial}{m^2} K^2\left(\frac{\partial\bar\partial}{m^2}\right)\right] \bar{\partial} J + \ldots.
\end{equation}
Therefore, we see that contribution of this term into $\cal R$ is of the form
$L K^2(L)$.

Up to this point our discussion was fairly straightforward and general. The difficult part however is the last $T_{KN} P$ term in (\ref{eq:PSchrodinger}).
There are several hard issues here, among which include the utility of the holomorphic invariant regulator, normal ordering and renormalization issues, and the choice of basis for
operators that we have taken (and consequent operator mixing). At the moment we do not have  full analytic control of these issuses. Therefore in what
follows we present some heuristic arguments to motivate our main {\it conjecture} in eq. (\ref{eq:TOn}). Further technical details can be found in Appendix~A.

The simplest way to think of the kinetic energy operator $T_{KN}$ (in particular the $J\delta/\delta J$ term) is
that it acts homogenously on any local operator-valued function of
$J$. According to this the action of $T_{KN}$ on $P$ from
Eq.(\ref{eq:Pquadratic}) would simply give us $2m P$ since $P$ is
quadratic in $\bar\partial J$. This picture, however, is certainly
not complete as can be easily seen by considering, for example, a
gauge and holomorphic invariant operator
\begin{equation}\label{eq:O1example}
{\cal O}_1 = \int \bar\partial J^a (D\bar\partial)^{ab} \bar\partial J^b = \int \bar\partial J^a (\partial\bar\partial)^{ab} \bar\partial J^a +
\int \bar\partial J^a \left(-i\frac{\pi}{c_A} f^{adb} J^d\bar\partial\right) \bar\partial J^b.
\end{equation}
$T_{KN}$ acting on this should give a holomorphically invariant. The $J\delta/\delta J$ term will act to count the number of $J$'s in each term -- this part of $T_{KN}$ is not holomorphic invariant, and thus the rest of $T_{KN}$ must act so as to restore holomorphic invariance. 
In particular, for ${\cal O}_1$ an extra quadratic in $J$ term is generated and we {\it expect} to obtain
\begin{equation}
T_{KN}\, {\cal O}_1 = 3m\, {\cal O}_1.
\end{equation}

Motivated by this heuristic argument\footnote{Other possible motivation for eq. (\ref{eq:TOn}) comes from comparison with lattice gauge theory.
We plan to elucidate this issue in a future publication.} we {\it expect} that the general result should be
\begin{equation}\label{eq:TOn}
T_{KN}\,\CO_n = (2+n)m\, \CO_n+\ldots
\end{equation}
where  the ellipsis stands for terms of higher order in $\bar\pa J$ but the same mass dimension as $\CO_n$ and which will mix with $\CO_n$ under the action
of $T_{KN}$. As we said above and would like to repeat here once again, we do not have complete analytic control of the action $T_{KN}$
and therefore at the moment  eq. (\ref{eq:TOn}) {\it cannot} be directly derived from the KKN Hamiltonian.  We will discuss these issues at some length in  Appendix~A. Nevertheless, as will be seen from what follows, (\ref{eq:TOn}) should be considered preliminary and conjectural, but its validity can in some sense be justified by the fact that it leads to sensible
physical results, which, without further apology, we explore  below.

Returning to the  derivation of the Schr\"odinger equation, asssuming (\ref{eq:TOn}) we can
write now
\beq
T_{KN}P=-\frac{\pi}{c_A m} \int \bar\pa J \left\{\frac{1}{2}\sum_n c_n(2+n)L^n \right\}\bar\pa J.
\eeq
It is convenient to write the factor in braces formally as
\begin{equation}
\frac{1}{2 L} \frac{d}{dL}\left[ L^2K(L)\right].
\end{equation}
Assembling all of these results, we then find
\begin{equation}\label{eq:Riccati}
{\cal R}= \frac{\pi}{c_A m}\left[-\frac{1}{2 L} \frac{d}{dL}\left[ L^2K(L)\right] + L K^2 +1\right],
\end{equation}
and if we set ${\cal R}$ to zero, we will finally obtain an equation for the kernel $K$
\begin{equation}\label{eq:RiccatiDiff}
-K - \frac{L}{2} \frac{d}{dL}[K(L)] + L K^2 +1 = 0.
\end{equation}

\subsection{ The Kernel}

Thus, we have succeeded in reducing the Schr\"odinger equation to a differential equation (\ref{eq:RiccatiDiff}) for the kernel, which is of the Riccati type. We will present complete details of its solution in Appendix~C, and note the important features here. Although the Riccati equation is non-linear, it is easily transformed into a linear second-order equation of the Bessel type, and one finds a general solution of the form
\begin{equation}
K(L)=\frac{1}{\sqrt{L}}\frac{CJ_2(4\sqrt{L})+Y_2(4\sqrt{L})}{CJ_1(4\sqrt{L})+Y_1(4\sqrt{L})}
\eeq
where $C$ is a constant and $J_n$ ($Y_n$) denote the Bessel functions of the first (second) kind.
As explained in Appendix~C, it is remarkable that {\it the only normalizable wave functional is obtained for $C\to \infty$, which is also the only case that has both the correct UV behavior appropriate to asymptotic freedom, as well as the correct IR behavior appropriate to confinement and a mass gap!}
This solution is of the form
\begin{equation}\label{eq:normvacuum}
K(L) = \frac{1}{\sqrt{L}} \frac{J_2(4 \sqrt{L})}{J_1(4 \sqrt{L})}.
\end{equation}
This remarkable formula is reminiscent of similar results in related
contexts \cite{migdal}; here, it encodes information on the spectrum of the theory, as
we show below. We note that this kernel has the following asymptotics
(where $L \sim -\vec p^2/4m^2$)
\begin{equation}
p \to 0, \quad K \to 1;\ \ \ \ \   p \to \infty, \quad K \to 2m/p
\end{equation}
consistent with confinement and asymptotic freedom, respectively.\footnote{Note that the argument of Bessel functions is imaginary;
so instead of $J_n$ we have $I_n$ Bessel functions.}

Now using standard Bessel function identities 
(see Appendix~C, Eqs.~(\ref{eq:besselrelns})) we may expand
\begin{equation}
\frac{J_{1}(u)}{J_{2}(u)} = \frac{4}{u} + 2 u \sum_{n = 1}^{\infty}\, \frac{1}{u^2 - \gamma_{2,n}^2}
\end{equation}
where the $\gamma_{2,n}$ are the ordered zeros of $J_2(u)$. For example, the first few zeros \cite{Bessel} of $J_2(u)$ are
$j_{2,1} = 5.14$, $j_{2,2} = 8.42$, $j_{2,3} = 11.62$, $j_{2,4} = 14.80$, {\it etc}.
The inverse kernel is thus
\begin{equation}
K^{-1}(L) = \sqrt{L}\,\frac{J_{1}(4 \sqrt{L})}{J_{2}(4 \sqrt{L})}
= 1 + 8L\sum_{n = 1}^{\infty}\,
 \frac{1}{16 L - \gamma_{2,n}^2}.
\end{equation}
Now, if we regard $L\simeq \pa\bar\pa /m^2$, in terms of momentum $\vec p$ we find
\begin{equation}\label{eq:KernelInverseK}
K^{-1}(p) =  1 + \frac{1}{2} \sum_{n = 1}^{\infty}\, \frac{\vec p^2}{\vec p^2 + M_{n}^2}.
\end{equation}
Here
\begin{equation}\label{constmass}
M_n = \frac{\gamma_{2,n} m}{2}.
\end{equation}
As we will see in the next section, $M_n$'s can be interpreted as constituents out of which glueball masses are constructed. The first
few of these $M_n$'s are
\begin{equation}\label{eq:constmasses}
\begin{array}{lcl}
M_1 & = 2.568 m\\
M_2 & = 4.209 m\\
M_3 & = 5.810 m\\
M_4 & = 7.398 m\\
M_5 & = 8.980 m.\\
\end{array}
\end{equation}
It is not difficult now to find a Fourier transform of inverse kernel $K^{-1}(p)$. By rewriting (\ref{eq:KernelInverseK}) as
\begin{equation}
K^{-1}(p) =  1 + \frac{1}{2} \sum_{n = 1}^{\infty}\, \left(1-\frac{M_{n}^2}{\vec p^2 + M_{n}^2}\right)
\end{equation}
we immediately obtain
\begin{equation}\label{eq:KFourier}
K^{-1}(x-y) = \delta^{(2)}(x-y) + \frac{1}{2} \sum_{n = 1}^{\infty}\, \left( \delta^{(2)}(x-y) - \frac{M_n^2}{2\pi}\, K_0(M_n |x-y|)  \right)
\end{equation}
where $K_0(x)$ is the modified Bessel function of the third kind.
At asymptotically large spacial separations $|x-y| \to \infty$
this takes the form
\begin{equation}\label{eq:Kpossp}
K^{-1}(|x-y|) \approx -\frac{1}{4\sqrt{2\pi |x-y|}} \sum_{n = 1}^{\infty} (M_{n})^{\frac{3}{2}}  e^{-M_n |x-y|} .
\end{equation}
This is a primary result, which will be important in our discussion of correlation functions in the next section. It is important to appreciate
that one should not think of the kernel $K$ as a propagator of some propagating field. Rather, as we discuss in
Appendix~D, it is a building block for such propagators.

Finally, to compute correlators at spatial separation, such as $\langle \bar\partial J^a(x)\ \bar\partial J^b(y)\rangle$, we have to evaluate a path integral over the
$H$ field with the insertion of the non-trivial WZW measure as well as $|\Psi_0^2|$. This would be a very difficult task unless we notice
that $\bar\partial J$ in fact can be treated as a free field. Below we present formal arguments supporting this observation; one may also consider this question formally based on a study of the path integral measure on the space of $J$'s, although we do not present this here.

We start with the observation that $J$ variable
can be obtained from $A$ by complex gauge transformation with matrix $M^{\dagger}$
\begin{equation}
A = -\partial M M^{-1}\quad \mapsto \quad A^{M^\dagger} = M^\dagger A M^{\dagger\, -1} - \partial M^\dagger M^{\dagger\, -1} \equiv -\frac{\pi}{c_A} J.
\end{equation}
Note also that under such transformation
\begin{equation}
\bar A = -\bar\partial M^{\dagger\, -1} M^\dagger \quad \mapsto \quad \bar A^{M^\dagger} = M^\dagger \bar A M^{\dagger\, -1} -
\bar\partial M^\dagger M^{\dagger\, -1} \equiv 0
\end{equation}
and  we may think of transformation from $A$, $\bar A$ variables
to $J$ variables as "gauge fixing" with gauge condition $\bar A =
0$. This is very similar to standard treatment of $2D$ Yang-Mills
theory \cite{Makeenko} in which case after the choice of some
axial gauge, say $A_1 = 0$, the action simply becomes
\begin{equation}
S_{2D} = \frac{1}{2 g_{2D}^2} \int \partial_1 A_2^a\ \partial_1 A_2^a
\end{equation}
and we can treat the only remaining variable $A_2$ as a free field. Given this similarity, it should be clear that we can treat $J$ as a free
field in the sense that
\begin{equation}
\langle \bar\partial J^a(x)\ \bar\partial J^b(y)\rangle = \delta^{ab}\frac{c_A m^2}{2\pi} K^{-1}(x-y).
\end{equation}
Now it is not difficult to derive an expression for the vacuum expectation of the large Wilson loop (\ref{eq:WilsonLoop}) of area $A$. To leading order
\begin{equation}
\langle \Phi(C)\rangle = \langle {\rm Tr}\, P\, {\rm exp}\left(\frac{\pi}{c_A}\oint_C J\right)  \rangle \ \ \longrightarrow \ \
N\, {\rm exp}\left( -\frac{N\pi^2}{c_A^2} \int_{A} \!\!\!{d^2x} {d^2y}\ \langle \bar\partial J(x)\ \bar\partial J(y)\rangle \right)
\end{equation}
or
\begin{equation}
\frac{1}{N} \langle\Phi(C)\rangle = {\rm exp} \left(-\frac{g_{YM}^4 (N^2-1)}{8\pi}\int d^2x d^2y\, K^{-1}(x-y)\right)
\end{equation}
and from this expression we see that the leading
$\delta$-function in (\ref{eq:KFourier}) gives an area law with the
same string tension (\ref{stringtension}) as was found before by
KKN. As for the rest of terms that appear in (\ref{eq:KFourier}),
we may notice that
\begin{equation}
\int_A d^2x  \left( \delta^{(2)}(x-y) - \frac{M_n^2}{2\pi}\, K_0(M_n |x-y|)  \right)\ \to\ 0 \quad {\rm as} \quad A\to\infty
\end{equation}
and therefore these terms will give corrections to the area law
behavior which vanish for asymptotically large loops. The
appearance of the $K_0$ Bessel functions suggests that our wave
functional secretely knows about some effective abelian vortex
configurations which model the repulsive two-particle part of the
ground state. This is familiar in strongly interacting systems in
condensed matter physics, where often the ground state of strongly
coupled systems is modelled as a product of single particle wave
functionals (the obvious WKB part of our wave functional) times an
effectively repulsive two-particle part which minimizes the energy
of the ground state by keeping the effective quasiparticles (the
$J$-``particles'') apart.

\section{The glueball spectrum and the Regge trajectories}

The primary purpose of this section is to provide a detailed
analysis of analytic computations of the glueball mass spectrum and to
compare these results with available lattice data \cite{teper}.
We will probe the spectrum of the theory by considering two point
functions of appropriate invariant operators at large spatial
separation. Such operators may be classified by their $J^{PC}$
quantum numbers, which we have described in detail in Section
\ref{sec:formalism}. For example, we will consider the $0^{++}$
states which may be probed by the operator ${\rm Tr}\,(\bar\pa
J\bar\pa J)$ and the corresponding correlator is
\begin{equation}\label{0++correlator}
\langle {{\rm Tr}\,(\bar{\partial} J \bar{\partial} J)}_x \
{{\rm Tr}\,(\bar{\partial} J \bar{\partial} J)}_y\rangle .
\end{equation}

But before continuing with
this analysis, let us make a few comments on this point. One expects that the glueballs are extended spatially in some way
(indeed, a common phenomenological picture is to consider the resonances as vibrational modes of a closed flux string).
We will derive the masses of these states by probing them with local operators. It is likely that this procedure has limited
applicability; in particular, masses of higher spin states are likely difficult to extract reliably. These limitations
also apply to the lattice work.  Therefore we will confine ourselves to a detailed analysis of spin-$0$ and spin-$2$ sectors
of the theory, and then indicate what the general picture might look like.

\subsection{Spin zero states}
\renewcommand{\thetable}{\arabic{table}}

We begin with the $0^{++}$ states. As we remarked above, we may
probe these states using the local operator ${\rm Tr}\, (\bar\pa
J\bar\pa J)$. Given our knowledge of the vacuum wavefunctional
(\ref{eq:psians}) we may compute the correlator
(\ref{0++correlator}) as
\begin{equation}
\langle {{\rm Tr}\,(\bar{\partial} J \bar{\partial} J)}_x \
{{\rm Tr}\,(\bar{\partial} J \bar{\partial} J)}_y\rangle \sim \left( K^{-1}(|x-y|)\right)^2
\end{equation}
This is of course not exact as we have dropped all interactions,\footnote{It would be of interest to consider the effect of interactions. For example, one would expect that these give rise to decay widths and mixings of states. Such an analysis would also likely clarify the identity and role of various expansions that underly our construction.} but we have kept the non-local spacetime dependence of the kernel.
Together with the expression for the inverse kernel (\ref{eq:Kpossp}) this gives
\begin{equation}\label{spin0correlator}
\langle {{\rm Tr}\,(\bar{\partial} J \bar{\partial} J)}_x \
{{\rm Tr}\,(\bar{\partial} J \bar{\partial} J)}_y\rangle
\approx
\frac{1}{32 \pi |x-y|} \sum_{n,\, m =1}^{\infty} (M_n M_m)^{3/2}e^{-(M_n+M_m)|x-y|}.
\end{equation}
From the characteristic exponential decay we can now read off the $0^{++}$ glueball masses:
\begin{equation}\label{masses}
\begin{array}{lcl}
M_{0^{++}} & =& M_1 + M_1 = 5.14 m\\
M_{0^{++*}} & =& M_1 + M_2 = 6.78 m\\
M_{0^{++**}} & =& M_1 + M_3 = 8.38 m\\
M_{0^{++***}} & =& M_1 + M_4 = 9.97 m\\
M_{0^{++****}} & =& M_1 + M_5 = 11.55 m.\\
\end{array}
\end{equation}
Since $m$ is not a physical scale, we should re-write these results in
terms of the string tension. Given equation (\ref{stringtension}) presented above, we can immediately
write for the lightest $0^{++}$ state 
\begin{equation}\label{ratio}
\frac{M_{0^{++}}}{\sqrt\sigma} = 5.14\,\sqrt{\frac{2}{\pi}}\,\frac{N}{\sqrt{N^2-1}}
\end{equation}
and similar expressions for the rest of glueballs listed in (\ref{masses}). Note, however, that equation (\ref{stringtension}) for the string
tension carries an explicit dependence on the rank $N$ of the guage group. Because of this, $M/\sqrt\sigma$ ratios depend explicitly on $N$ as well.
As an example, in Table \ref{Table01} comparison of our predictions for the mass of the lightest $0^{++}$ state at different values of $N$ ($N=2,3,4,5,6,\infty$) with
lattice data is given. From this we see that simple "Casimir scaling" encoded in (\ref{ratio}) describes $N$-dependence extremelly well even for $N$ as low as $N = 2$.
For this reason in the rest of our discussion of glueball mass spectrum, we will limit ourselves to comparison with lattice at $N=\infty$ only. 
Similar comparisons at finite $N$ should present no difficulties. In particular, our results for $0^{++}$ states at $N=\infty$ are given in Table \ref{Table1}.

\begin{table}\centering
\caption{Dependence of the mass of the lightest $0^{++}$ glueball on the rank of the gauge group. All masses
are in units of the square root of the string tension. The percent difference between our predictions and lattice data is given
in the last column.}\label{Table01}
\bigskip
\begin{tabular}{c|ccc}
\hline\hline
Gauge group & Lattice & Our prediction & Diff, \% \\
\hline
$SU(2)$ & $4.716 \pm 0.021$ & $4.732$ & $0.3$ \\
$SU(3)$ & $4.330 \pm 0.024$ & $4.347$ & $0.4$\\
$SU(4)$ & $4.235 \pm 0.025$ & $4.233$ & $0.05$\\
$SU(5)$ & $4.180 \pm 0.039$ & $4.182$ & $0.05$\\
$SU(6)$ & $4.196 \pm 0.027$ & $4.156$ & $1.0$\\
$SU(\infty)$ & $4.065 \pm 0.055$ & $4.098$ & $0.8$ \\
\hline\hline
\end{tabular}
\end{table}
 
\begin{table}\centering
\caption{$0^{++}$ glueball masses in $QCD_3$. All masses
are in units of the square root of the string tension. Results of $AdS/CFT$
computations in the supergravity limit are also given for comparison.
The percent difference between our prediction and lattice data is given
in the last column.}\label{Table1}
\bigskip
\begin{tabular}{l|cccc}
\hline\hline
State & Lattice, $N\to\infty$ & Sugra & Our prediction & Diff, \% \\
\hline
$0^{++}$ & $4.065 \pm 0.055$ & $4.07$(input) & $4.098$ & $0.8$ \\
$0^{++*}$ & $6.18 \pm 0.13$ & $7.02$ & $5.407$ & $12.5$\\
$0^{++**}$ & $7.99 \pm 0.22$ & $9.92$ & $6.716$ & $16$\\
$0^{++***}$ & $9.44 \pm 0.38$
\footnotemark{}
& $12.80$ & $7.994$ & $15$\\
$0^{++****}$ & $--$ & $15.67$ & $9.214$ & $--$\\
\hline\hline
\end{tabular}
\end{table}
\footnotetext{Mass of $0^{++***}$ state was
computed on the lattice for $SU(2)$ only~\cite{Meyer:2003wx}.
The number quoted here was obtained by a simple rescaling of $SU(2)$
result.}

Several comments are now in order. First, there are not any
adjustable parameters in our theory; the ratio of $M_{0^{++}}$ to
$\sqrt\sigma$ {\it is a pure number} and is in excellent agreement
with lattice data! This should be contrasted with the supergravity
results also listed in the table for comparison  and which result
from calculations \cite{Terning} based on the AdS/CFT
correspondence \cite{adscft}; in that case, the overall
normalization {\it is not} predicted but is instead determined by
fitting to the lattice data, for example, to the mass of the
lowest $0^{++}$ glueball. Second, note that we have been able to
predict masses of the $0^{++}$ resonances, as well as the lowest
lying member. Our
results for the excited state masses differ at the 10-15\% level
from the lattice results. We note that precisely these masses are
more difficult to compute on the lattice \cite{OZ}, and thus the
apparent $10-15\%$ discrepancy may be illusory.

 Another possible explanation of such discrepancy comes
from the  observation that lattice values for the masses of $0^{++**}$ and
$0^{++***}$ states  are in
much better agreement with our prediction for $0^{++***}$ and $0^{++****}$ states respectively
({\it cf.}~$7.99~\pm~0.22$~{\it vs.}~$7.994$ and $9.44~\pm~0.38$~{\it vs.}~$9.214$.). Also,
 numerically, lattice prediction for  $0^{++*}$ mass seems to lie in between our numbers for
$0^{++*}$ and  $0^{++**}$ states thus indicating that it might be
some kind of an ``average'' . These two observations suggest the
following picture: the first two excited states have not been
properly resolved in lattice computations and have been counted as
a single mass eigenstate ``$0^{++*}$'' with an ``average'' mass of
$6.18 \pm 0.13 \sqrt\sigma$; as a consequence, the rest of the
excitations should be labelled differently and lattice
``$0^{++**}$'' state should be labelled as $0^{++***}$, while
lattice ``$0^{++***}$'' state should be labelled as $0^{++****}$.
Assuming this picture to be correct, we give an updated comparison
of  our $0^{++}$ mass predictions with lattice data in
Table~\ref{Table2}. Obviously, the numerical agreement is quite
impressive.

\begin{table}\centering
\caption{ Same as Table \ref{Table1} but with
lattice data relabelled as explained in text.}\label{Table2}
\bigskip
\begin{tabular}[b]{l|ccc}
\hline\hline
State & Lattice, $N\to\infty$ & Our prediction & Diff, \% \\
\hline
$0^{++}$ & $4.065 \pm 0.055$ & $4.098$ & $0.8$ \\
$0^{++*}$ & $6.18 \pm 0.13$ & $5.407$ & $--$\\
$0^{++**}$ & $6.18 \pm 0.13$ & $6.716$ & $--$\\
$0^{++***}$ & $7.99 \pm 0.22$ & $7.994$ & $0.05$\\
$0^{++****}$ & $9.44 \pm 0.38$ & $9.214$ & $2.4$\\
\hline\hline
\end{tabular}
\end{table}

Let us move on to a discussion of the $0^{--}$ glueball resonances.  We may probe these
states with the operator ${\rm Tr}\,(\bar{\partial} J \bar{\partial} J
\bar{\partial} J)$ .
We are thus interested in the correlation function
\begin{equation}
\langle {{\rm Tr}\,(\bar{\partial} J \bar{\partial} J \bar{\partial} J)}_x \ {{\rm Tr}\,(\bar{\partial} J \bar{\partial} J \bar{\partial} J)}_y\rangle
\sim \left( K^{-1}(|x-y|)\right)^3
\end{equation}
or
\begin{equation}\label{0--correlator}
\langle {{\rm Tr}\,(\bar{\partial} J \bar{\partial} J \bar{\partial} J)}_x \ {{\rm Tr}\,(\bar{\partial} J \bar{\partial} J \bar{\partial} J)}_y\rangle
\sim
\frac{1}{64(2 \pi |x-y|)^{3/2}} \sum_{n,\, m,\, k =1}^{\infty} (M_n M_m M_k)^{3/2}e^{-(M_n+M_m+M_k)|x-y|}.
\end{equation}
Using this result, we obtain $0^{--}$ glueball masses which are the
sum of three $M_n$'s
\begin{equation}
\begin{array}{ll}
M_{0^{--}} & =  M_1 + M_1 + M_1  =  7.70 m \\
M_{0^{--*}} & =  M_1 + M_1 + M_2  =  9.34 m \\
M_{0^{--**}} & =  M_1 + M_1 + M_3  =  10.95 m. \\
\end{array}
\end{equation}
These results are compared to lattice and supergravity data in Table \ref{Table3}.
We see that our predictions are within a few percent of the lattice
data, and are much better than the supergravity predictions.

One last comment about $0^{--}$ correlator in (\ref{0--correlator}). Although 
it has
an exponentially falling nature, the prefactor is not of the correct form so as to be
directly interpreted as a single particle pole (for details, see Appendix~D.) We believe that this discrepancy can
be accounted for by normalizing the probe operator appropriately. Note that this issue
does not arise for the $0^{++}$ states.

\begin{table}\centering
\caption{$0^{--}$ glueball masses in $QCD_3$. All masses
are in units of the square root of the string tension. Results of $ADS/CFT$
computations in the supergravity limit are also given for comparison.
The percent difference between our prediction and lattice data is given
in the last column.}\label{Table3}
\bigskip
\begin{tabular}{l|cccc}
\hline\hline
State & Lattice, $N\to\infty$ & Sugra & Our prediction & Diff,\%\\
\hline
$0^{--}$ & $5.91 \pm 0.25$ & $6.10$ & $6.15$ & $4$ \\
$0^{--*}$ & $7.63 \pm 0.37$ & $9.34$ & $7.46$ & $2.3$ \\
$0^{--**}$ & $8.96 \pm 0.65$ & $12.37$ & $8.73$ & $2.5$ \\
\hline\hline
\end{tabular}
\end{table}

\subsection {Spin $2$ states}

As we have seen in the previous subsection, there is a fairly good
amount of ``experimental'' lattice data available for spin-$0$
sector of $2+1$ dimensional pure Yang-Mills theory. Also the
quality of those data appears to be good enough thus making direct
comparison with our predictions possible. Unfortunately this is
not the case with higher spin states: in most cases only the mass
of the lowest resonance with given $J^{PC}$ quantum numbers is
known (at best) and this, combined with much larger
``experimental'' (lattice)  error bars, makes for a much more
problematic comparison. The only exceptions are spin-$2$ states
and it is our intention to discuss them next. However, given the
comments presented above regarding the amount and quality of the
lattice data available at the moment of the writing of this paper,
one should not really expect such an excellent agreement as we had
in spin-$0$ sector (although,  as will be seen below, the
agreement is still quite impressive).

But before we proceed any further let us remind the reader about one peculiar property of $2+1$ dimensional Yang-Mills theory
known as {\it parity doubling}. Consider a state $|P,J; E\rangle$ of certain parity $P$, angular momentum $J$ and energy $E$. Note however
that since parity and angular momentum operators do not commute (they rather anticommute, $\{{\hat J}, {\hat P}\}=0$, meaning that parity flips sign under the
action of $\hat J$) this state is not an eigenstate of
$\hat J$ but of $\hat J^2$ only. Therefore by acting on this state with ${\hat J}$ we obtain an essentially different state (or zero if $J=0$)
\begin{equation}
|-P,J; E\rangle = {\hat J} |P,J; E\rangle
\end{equation}
of the same energy $E$, angular momentum $J$ but opposite parity. This means in particular that masses of glueballs with the same $J$ and $C$
quantum numbers but opposite parities should be the same. Of course, this argument does not apply to $J=0$ states and there is no reason to
expect parity doubling in the spin zero sector of the theory.

This phenomenon of parity doubling simplifies our task since all we have to do is to find masses of $2^{++}$ and $2^{--}$ resonances; then these
results should apply, respectively, to $2^{-+}$ and $2^{+-}$ states as well. For $2^{++}$ states we may take the simplest possible spin-$2$
operator namely ${\rm Tr}\, ({\bar\pa}^2 J{\bar\pa}^2 J)$ and compute the correlator
\begin{equation}\label{spin2correlator}
\langle {{\rm Tr}\, ({\bar\pa}^2 J{\bar\pa}^2 J)}_x \ {{\rm Tr}\, ({\bar\pa}^2 J{\bar\pa}^2 J)}_y \rangle \sim
 \left( \bar\partial_x \bar\partial_y K^{-1}(|x-y|)\right)^2
\end{equation}
or
\begin{equation}
\langle {{\rm Tr}\, ({\bar\pa}^2 J{\bar\pa}^2 J)}_x \ {{\rm Tr}\, ({\bar\pa}^2 J{\bar\pa}^2 J)}_y \rangle \sim
\frac{|x-y|^3}{2\pi 4^4 (\bar x - \bar y)^4} \sum_{n,\, m =1}^{\infty} (M_n M_m)^{7/2}e^{-(M_n+M_m)|x-y|}
\end{equation}
to obtain a paradoxical result: masses of $2^{++}$ states are given by the sums of two constituent masses $M_n$, {\it i.e} are the same as masses of
$0^{++}$ states! The resolution of this paradox lies apparently in the fact that we use local operators to probe extended objects. As a result,
the correlator in (\ref{spin2correlator}) (or (\ref{spin0correlator})) describes not only spin-$2$ (or spin-$0$) resonances but states of other spins as well.
Going back to our discussion of $0^{++}$ sector, one may
notice that in (\ref{masses}) we identified as $0^{++}$ only those states with one of the constituent masses equal to $M_1$ (i.e. these states are of the form
$M_1+ M_n,\ n=1,2,\ldots$). It should now be clear that $2^{++}$ states are described by the series with  one of the constituent masses equal to $M_2$,
 $4^{++}$ states - by series with one $M_3$, {\it etc} (we will say more on higher spin states in the next subsection).

\begin{table}\centering
\caption{$2^{\pm+}$ glueball masses in $QCD_3$.
All masses are in units of the square root of the string tension.}\label{Table4}
\bigskip
\begin{tabular}{l|ccc}
\hline\hline
State & Lattice, $N\to\infty$ & Our prediction & Difference, \%\\
\hline
$2^{++}$ & $6.88 \pm 0.16$ &  $6.72$ & $2.4$\\
$2^{-+}$ & $6.89 \pm 0.21$ &  $6.72$ & $2.5$\\
$2^{++*}$ & $8.62 \pm 0.38$ &  $7.99$ & $7.6$\\
$2^{-+*}$ & $9.22 \pm 0.32$ &  $7.99$ & $14$\\
$2^{++**}$ & $10.6 \pm 0.7$
\footnotemark{}
&  $9.26$ & $13$\\
$2^{++***}$ & $--$ &  $10.52$ & $--$\\
\hline\hline
\end{tabular}
\end{table}
\footnotetext{Mass of $2^{++**}$ state was
computed on the lattice for $SU(8)$ only~\cite{Teperprivate}.
The number quoted here was obtained by a simple rescaling of $SU(8)$
result.}

Thus our prediction for the masses of $2^{++}$ (as well as $2^{-+}$) states is:
\begin{equation}\label{2++masstable}
\begin{array}{ll}
M_{2^{++}} & =  M_2 + M_2 =  8.42 m \\
M_{2^{++*}} & =  M_2 + M_3  =  10.02 m \\
M_{2^{++**}} & =  M_2 + M_4  =  11.61 m \\
M_{2^{++***}} & =  M_2 + M_5  =  13.19 m. \\
\end{array}
\end{equation}
In Table \ref{Table4} this data is compared with the available
results of lattice calculations. We see that like in $0^{++}$ case
we have excellent agreement for the lowest resonance while for
excited states, the discrepancy is again at the $10-15\%$ level. Let us
note however that the lattice values for $2^{++*}$ and $2^{-+*}$
states are quite different numerically (even though formally they
are within the error bars) and this allows us to question their
validity. This can be compared, for example, to the lattice
predictions for $2^{++}$ and $2^{-+}$ states which are almost
identical (as they should be, due to parity doubling)  and so seem
to be reliable. Furthermore, as in the $0^{++}$ case,  we notice
that the lattice value for $2^{-+*}$ state is numerically much
closer to our prediction for the $2^{++**}$ state, while the
lattice value for $2^{++**}$ state is close to our prediction for the mass of the $2^{++***}$
state. Therefore, it is interesting to consider the possibility
that lattice states have not been properly identified and that
proper labelling should be as follows: the lattice mass of the
$2^{++*}$ state can probably be considered as a reasonable
approximation to the mass of a true $2^{++*}$ state, but the
lattice $2^{-+*}$ state should be identified as the true
$2^{++**}$ state. Finally, the lattice $2^{++**}$ state should be
compared to our $2^{++***}$ state. The comparison of our analytic
predictions with corresponding lattice results after such
relabelling is presented in Table \ref{Table5}.  Once again, the
numerical agreement is excellent.

Let us proceed now to $2^{--}$ states. For these states we may take ${\rm Tr}\, (\bar\pa J{\bar\pa}^2 J {\bar\pa}^2 J)$ as the relevant probe
operator. By computing the correlator of two such probe operators as we did before, we obtain that the masses of $2^{--}$ resonances as sums of three
$M_n$'s, i.e. formally this is the same result as we have previously found for $0^{--}$ states. However by the same kind of reasoning as we had in the $2^{++}$ {\it vs.} $0^{++}$ case, we
believe that the mass series for $2^{--}$ resonances should begin with $M_1 + M_2 + M_2$. Thus our prediction for $2^{--}$ sector of the theory is
\begin{equation}
\begin{array}{ll}
M_{2^{--}} & =  M_1 + M_2 + M_2 =  10.99 m \\
M_{2^{--*}} & =  M_1 + M_2 + M_3  =  12.59 m \\
M_{2^{--**}} & =  M_1 + M_2 + M_4  =  14.18 m .\\
\end{array}
\end{equation}

For these states, there are not that many available lattice data. We list these together with our predictions
in Table \ref{Table6}. Again we can see that agreement is reasonably good especially given the much larger error bars for
the spin-$2$ lattice data as compared
to their spin zero counterparts.

\begin{table}\centering
\caption{Same as Table \ref{Table4} but with lattice data relabelled as explained in text.}\label{Table5}
\bigskip
\begin{tabular}{l|ccc}
\hline\hline
State & Lattice, $N\to\infty$ & Our prediction & Difference, \%\\
\hline
$2^{++}$ & $6.88 \pm 0.16$ &  $6.72$ & $2.4$\\
$2^{++*}$ & $8.62 \pm 0.38$ &  $7.99$ & $7.6$\\
$2^{++**}$ & $9.22 \pm 0.32$ &  $9.26$ & $0.4$\\
$2^{++***}$ & $10.6 \pm 0.7$ &  $10.52$ & $0.8$\\
\hline\hline
\end{tabular}
\end{table}

\begin{table}\centering
\caption{$2^{\pm-}$ glueball masses in $QCD_3$. All masses are in units of the square root of the string tension.}\label{Table6}
\bigskip
\begin{tabular}{l|cccc}
\hline\hline
State & Lattice, $N\to\infty$ & Our prediction & Difference, \%\\
\hline
$2^{+-}$ & $8.04 \pm 0.50$ & $8.76$ & $8.6$\\
$2^{--}$ & $7.89 \pm 0.35$ & $8.76$ & $10.4$\\
$2^{+-*}$ & $9.97 \pm 0.91$ & $10.04$ & $0.7$\\
$2^{--*}$ & $9.46 \pm 0.66$ & $10.04$ & $5.6$\\
\hline\hline
\end{tabular}
\end{table}

\subsection{Higher spin states and Regge trajectories}

Unfortunately, at the moment, there are no lattice data available
for glueballs with spins higher than $2$ and it is our hope that
future lattice simulations will address this issue. However, based
on the spin-$0$ and spin-$2$ cases discussed above, it is possible
to envisage natural higher spin generalizations of our results. One might expect that the quasi-Gaussian approximation to the vacuum wavefunctional begins to become insufficient to account for the masses of higher spin states. In the context of this paper, we have no way to test this, and simply provide the reader with the predictions. In
particular, the masses of states with even spin and $C=+$ are
given by the sums of two constituent masses $M_k$ according to the
following rule ($n = 0,1,2,\ldots$):
\begin{itemize}
\item $0^{++}$ and corresponding resonances: $M_{0^{++*^n}}= M_1 + M_{1+n}$;
\item $2^{++}$ and corresponding resonances: $M_{2^{++*^n}}= M_2 + M_{2+n}$;
\item $4^{++}$ and corresponding resonances: $M_{4^{++*^n}}= M_3 + M_{3+n}$;
\item $6^{++}$ and corresponding resonances: $M_{6^{++*^n}}= M_4 + M_{4+n}$;
\end{itemize}
or, in general
\begin{equation}\label{J++mass}
M_{J^{++*^n}} = M_{J/2 + 1} + M_{J/2 + 1 + n}, \ \ \ \ \ J =0,2,4,6,\ldots .
\end{equation}
Here, for convenience, we choose to write these expressions for positive parity states only. One should remember that,
by parity doubling, the masses of corresponding negative parity states are identical (unless $J=0$).

Similarly, for the even spin resonances with $C=-$ we may write:
\begin{itemize}
\item $0^{--}$ and corresponding resonances: $M_{0^{--*^n}}= M_1 + M_1 + M_{1+n}$;
\item $2^{--}$ and corresponding resonances: $M_{2^{--*^n}}= M_1 + M_2 + M_{2+n}$;
\item $4^{--}$ and corresponding resonances: $M_{4^{--*^n}}= M_1 + M_3 + M_{3+n}$;
\item $6^{--}$ and corresponding resonances: $M_{6^{--*^n}}= M_1 + M_4 + M_{4+n}$;
\end{itemize}
and, in general
\begin{equation}\label{J--mass}
M_{J^{--*^n}} = M_1 + M_{J/2+1} + M_{J/2+1+n}, \ \ \ \ \ J =0,2,4,6,\ldots .
\end{equation}
Once again we write these expressions for $P=-$ states; the $P=+$ results are identical.
A few explicit examples of how Eqs.~(\ref{J++mass}, \ref{J--mass}) work are given in Table \ref{Table7}.

Given these generalized formulae for the masses of higher spin
states, we can represent them graphically in the form of a
Chew-Frautschi plot ($J$ {\it vs.} $M^2/\sigma$) and attempt to
identify Regge trajectories. This is done in Figures $1$ and $2$.
It should be noted, however, that to the best of our knowledge,
there is no fundamental reason for the existence of linear Regge
trajectories in the glueball sector of Yang-Mills theory.
Nevertheless, our plots indicate that it is still possible to draw
nearly linear Regge trajectories.
\begin{table}\centering
\caption{Higher spin ($4^{++}, 4^{--}, 6^{++}, 6^{--}, 8^{++}$) glueball masses in $QCD_3$. All masses are in units of the square root of the string tension.}\label{Table7}
\bigskip
\begin{tabular}{l|c}
\hline\hline
State  & Our prediction\\
\hline
$4^{++}$ &  $9.27$\\
$4^{++*}$ &  $10.54$\\
$4^{++**}$ &  $11.80$\\
\hline
$6^{++}$ &  $11.81$\\
$6^{++*}$ &  $13.07$\\
$6^{++**}$ &  $14.32$\\
\hline
$8^{++}$ &  $14.33$\\
$8^{++*}$ &  $15.59$\\
$8^{++**}$ &  $16.85$\\
\hline
\hline
$4^{--}$ &  $11.32$\\
$4^{--*}$ &  $12.59$\\
\hline
$6^{--}$ &  $13.86$\\
$6^{--*}$ &  $15.12$\\
\hline\hline
\end{tabular}
\end{table}

We would like to conclude this section by pointing out that our
discussion of the mass spectrum of $2+1$ dimensional pure
Yang-Mills theory is certainly not complete. In particular,
current computational scheme based on the use of local probe
operators  is well suited for glueballs with $PC = ++$ or $PC =
--$. However for resonances with $PC = -+$ or $PC = +-$ any probe
operator with such quantum numbers is essentially nonlocal and it
is not clear at the moment how our results generalize in that
case. We were able to appeal to parity doubling phenomenon to
circumvent this difficulty for states of nonzero spin $J$
(although it is still interesting to compute masses of $J^{+-}$
and $J^{-+}$ directly), but for $0^{-+}$ and $0^{+-}$ we cannot
do even that. It should be noted nevertheless, that it is still
possible to pick certain combinations of constituent masses $M_n$
to approximate existing lattice data on $0^{-+}$ and $0^{+-}$, but
theoretical reasons behind this are not clear.

Also we do not discuss here states of odd spin. However we would like to
point out that existing lattice data for the lowest $1^{\pm+}$ states
\begin{equation}
\begin{array}{ll}
M_{1^{++}} & =  9.98\pm 0.5\, \sqrt{\sigma} \\
M_{1^{-+}} & =  10.06\pm 0.4\, \sqrt{\sigma} \\
\end{array}
\end{equation}
can be approximated (up to $5\%$) by
\begin{equation}
M_{1^{++}}  = M_1+M_1+M_1+M_2 =  9.50\, \sqrt{\sigma} 
\end{equation}
while lattice values for $1^{\pm-}$
\begin{equation}
\begin{array}{ll}
M_{1^{+-}} & =  9.43\pm 0.75\, \sqrt{\sigma} \\
M_{1^{--}} & =  9.36\pm 0.60\, \sqrt{\sigma} \\
\end{array}
\end{equation}
should presumably correspond to
\begin{equation}
M_{1^{+-}}  = M_1+M_1+M_1+M_1+M_2 =  11.55\, \sqrt{\sigma}. 
\end{equation}
As a possible explanation of such a large difference (about $20\%$) in the last case we may notice that in general it seems that $C=-$ states are heavier
than $C=+$ states of the same spin ({\it cf}. $2^{++}$ and $2^{+-}$, for example). This is not the case with $1^{\pm-}$ lattice
data and therefore we may question its validity.  

\myfig{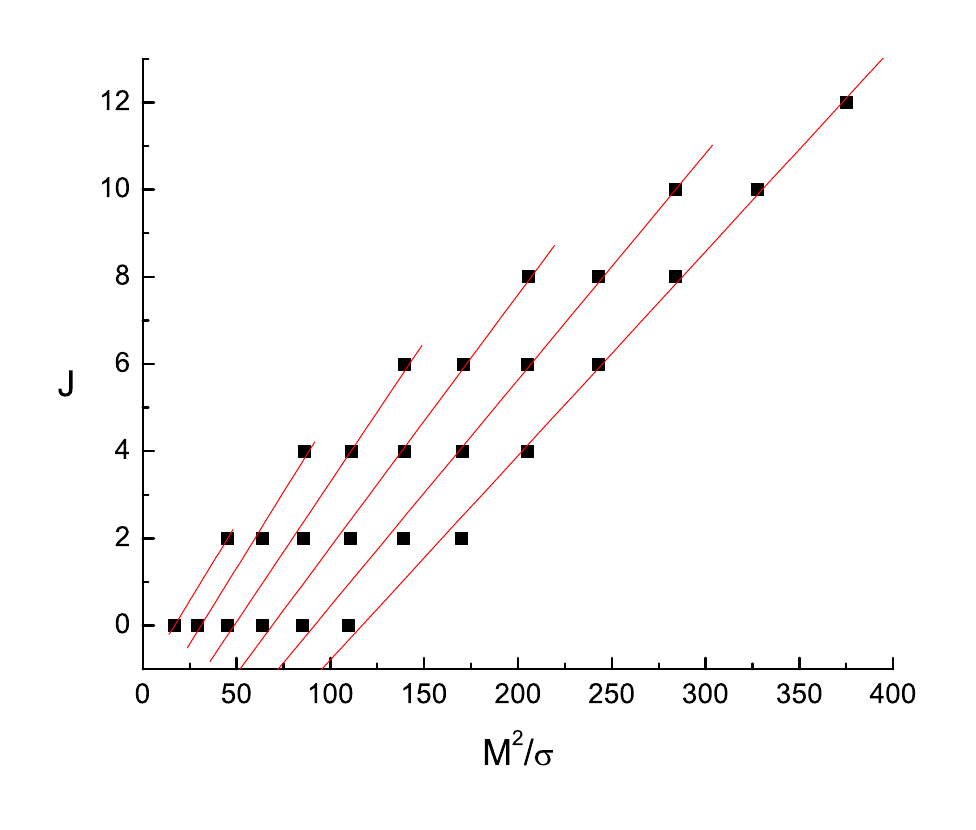}{9}{Chew-Frautschi plot of the large $N$ glueball spectrum. Black boxes correspond to $J^{++}$ glueball resonances
with even spins up to $J=12$. Solid red lines represent linear Regge trajectories.}

\myfig{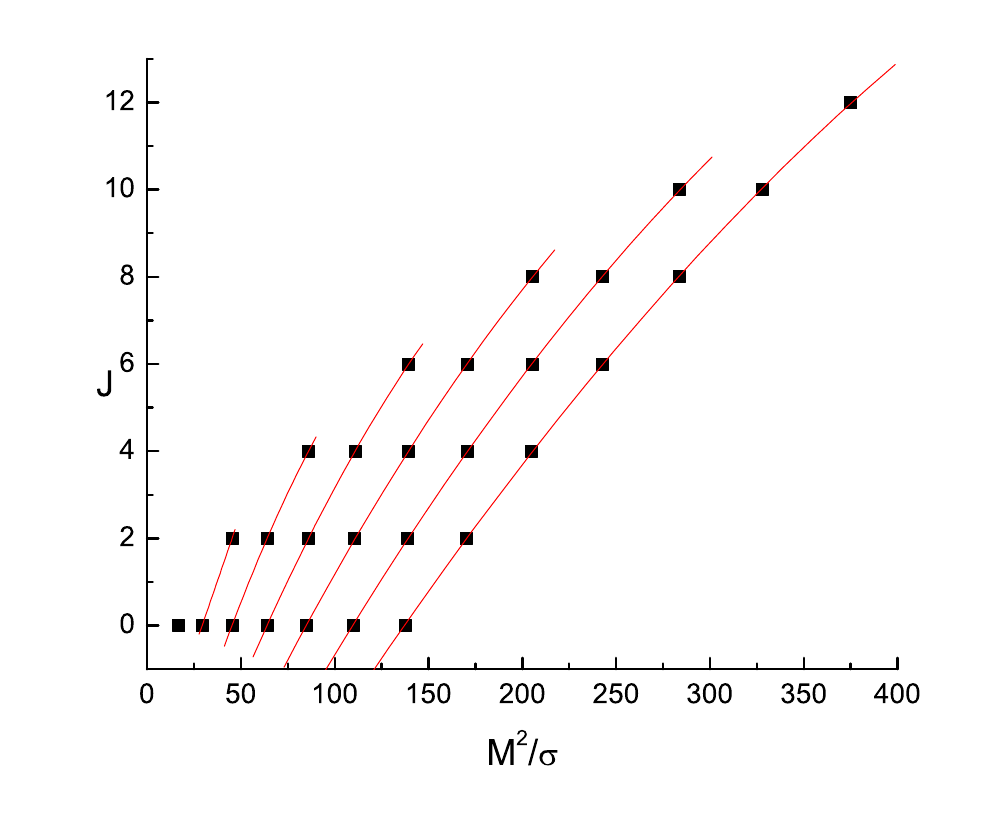}{9}{Same data as in Figure $1$. Red lines represent polynomial data fit with second order polynomial
$J = \alpha_0 + \alpha_1 \left(\frac{M^2}{\sigma}\right) + \alpha_2 {\left(\frac{M^2}{\sigma}\right)}^2$. Note also
the different way to connect data points compared to Figure $1$.}

\subsection{Fine Structure and Asymptotic Properties of Mass Spectrum }

As was extensively discussed in previous sections, glueball masses in $2+1$-dimensional pure Yang-Mills theory are
given by expressions involving zeros $\gamma_{2,i}$ of Bessel function $J_2 (x)$. For example, equation (\ref{J++mass}) describing
${J^{++*^n}}$ (for even $J$) resonances can be rewritten in terms of $\gamma_{2,i}$ as
\begin{equation}
M_{J^{++*^n}} = \frac{m}{2}\left(\gamma_{2,J/2+1} + \gamma_{2,J/2+1+n}\right).
\end{equation}

We can use now the following asymptotic expressions for the zeros of Bessel functions \cite{Watson}
\begin{equation}
\gamma_{\nu, n} = \pi \left(n + \frac{1}{2} \nu - \frac{1}{4}\right) - \frac{4\nu^2-1}{8\pi\left(n+\frac{1}{2}\nu - \frac{1}{4}\right)} + {\cal O}\left(\frac{1}{n^3}\right)
\end{equation}
\begin{equation}\label{Besselasymptotic}
\gamma_{2, n} = \pi \left(n + \frac{3}{4}\right) - \frac{15}{8\pi\left(n+\frac{3}{4}\right)} + {\cal O}\left(\frac{1}{n^3}\right)
\end{equation}
to obtain asymptotic expressions for highly excited $J^{++*^n}$states (with even $J$ and large $n$):
\begin{itemize}
\item[]
\begin{equation*}
M_{0^{++*^n}} = m\frac{\pi}{2}\left(n+\frac{7}{4} + \frac{\gamma_{2,1}}{\pi}\right) + {\cal O}\left(\frac{1}{n}\right)
\end{equation*}
\item[]
\begin{equation*}
M_{2^{++*^n}} = m\frac{\pi}{2}\left(n+\frac{11}{4} + \frac{\gamma_{2,2}}{\pi}\right) + {\cal O}\left(\frac{1}{n}\right)
\end{equation*}
\item[]
\begin{equation*}
\ldots\ldots\ldots\ldots\ldots\ldots\ldots\ldots
\end{equation*}
\item[]
\begin{equation}\label{asymptotic}
M_{J^{++*^n}} = m\frac{\pi}{2}\left(n+\frac{2J+7}{4} + \frac{\gamma_{2,J/2+1}}{\pi}\right) + {\cal O}\left(\frac{1}{n}\right).
\end{equation}
\end{itemize}
We can see now that for large excitation level $n$, the glueball
mass spectrum with given $J^{PC}$ quantum numbers becomes
equidistant with mass splittings 
\begin{equation}
\Delta M = M_{J^{++*^{(n+1)}}} - M_{J^{++*^n}} = m\frac{\pi}{2}.
\end{equation}
For large values of spin $J$ we can further simplify Eq.(\ref{asymptotic}) if we replace $\gamma_{2,J/2+1}$ by its asymptotic value. We get
\begin{equation}
M_{J^{++*^n}} = m\frac{\pi}{2}\left(n+J + \frac{7}{2}\right) + {\cal O}\left(\frac{1}{n}, \frac{1}{J}\right).
\end{equation}
This is an interesting expression since it tells us that there is approximate mass degeneracy in the theory. By this we mean that the mass
of a state with given large values of $n$ and $J$ is approximately the same as masses of states with
the corresponding excitation number and spin equal to $n+i$ and
$J-i$ respectively ($i = \pm 2, \pm 4, \ldots$). Actually, since Bessel zeros approach their asymptotic values very fast we can see such approximate
degeneracy even for states of low spin $J$. For example, as was found in (\ref{masses}) and (\ref{2++masstable}) masses of $2^{++}$ and $0^{++**}$ states
\begin{equation}
\begin{array}{ll}
M_{0^{++**}} & = M_1 + M_3 = 8.38 m\\
M_{2^{++}} & =  M_2 + M_2 =  8.42 m \\
\end{array}
\end{equation}
are very close numerically, the difference being about $.047\%$. Similarly for
\begin{equation}
\begin{array}{ll}
M_{0^{++***}} & = M_1 + M_4 = 9.97 m\\
M_{2^{++*}} & =  M_2 + M_3  =  10.02 m \\
\end{array}
\end{equation}
the difference is about $0.52\%$.

\myfig{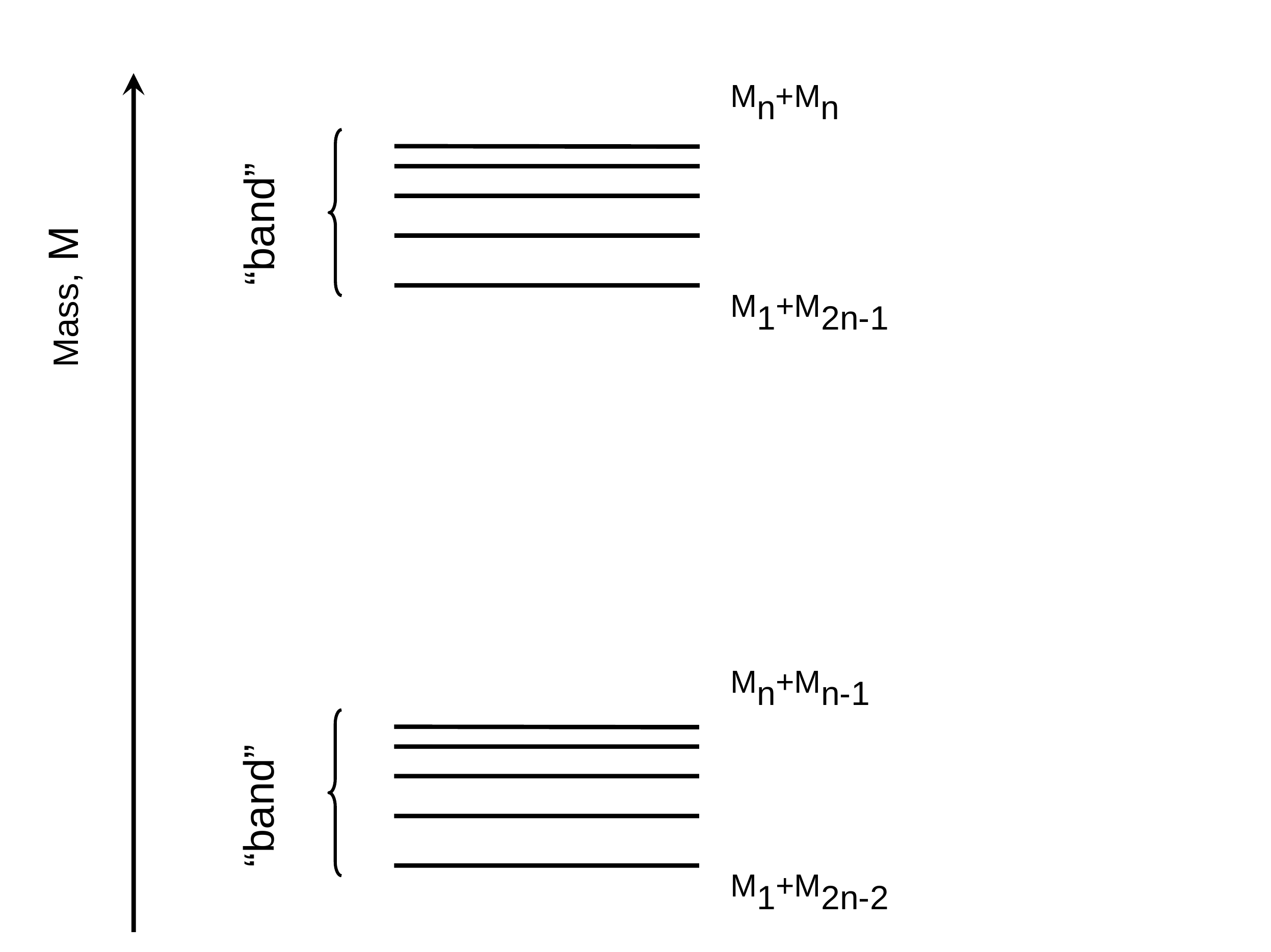}{9}{``Band'' structure of the mass spectrum.}

We thus see that the mass spectrum of the theory has a ``fine structure'' with numbers of minutely separated mass
eigenstates grouped together into ``bands''. 
It is interesting nevertheless to investigate their structure a bit further. In general, $2n$'s ``band'' is made out of
the states with the following mass content: $M_n + M_n$, $M_{n-1} + M_{n+1}$, $\ldots$, $M_1 + M_{2n-1}$ (and similarly for
``bands'' with odd number $2n+1$). To see their general features, it is convenient to work out an explicit example; say, for
the $50$th ``band'' we have
\begin{equation}\label{bandmasses}
\begin{array}{c}
M_1+M_{49} = 80.70893 m\\
M_2+M_{48} = 80.77882 m\\
\ldots\ldots\ldots\ldots\ldots\ldots\ldots\\
M_{23}+M_{27} = 80.872686 m\\
M_{24}+M_{26} = 80.872792 m\\
M_{25}+M_{25} = 80.872827 m.\\
\end{array}
\end{equation}
From this example it is not difficult to see that the heaviest
state within a ``band'' is the state with the most equal masses
($M_{25}+M_{25}$ in this example), while the lightest state is the
one with the most different constituent masses ($M_1+M_{49}$). The
``bands'' do not overlap: say, the most massive state in the
$49$th ``band'' is $M_{24}+M_{25} = 79.30155 m$ and so the
distance between two ``bands'' (about $1.5 m$) is about $10$ times
the width of the ``band'' (about $0.16 m$). Finally, as can be
seen from (\ref{bandmasses}), mass levels are very dense near the
top of the ``band'' and spread further apart towards the bottom.
Using Eq.(\ref{Besselasymptotic}) it is possible to derive the
needed asymptotic expression for the mass splittings within a
``band''. The highest level in the $2n$'th ``band'' is
\begin{equation}
M_n + M_n = \frac{m}{2}\left[2\pi\left(n+\frac{3}{4}\right) - \frac{30}{8\pi\left(n+\frac{3}{4}\right)}+ {\cal O}\left(\frac{1}{n^3}\right)\right].
\end{equation}
Next levels are
\begin{equation}
M_{n+\alpha} + M_{n-\alpha} =\frac{m}{2}\left[2\pi\left(n+\frac{3}{4}\right) - \frac{30}{8\pi\left(n+\frac{3}{4}\right)}
 - \frac{30}{8\pi\left(n+\frac{3}{4}\right)^3} \alpha^2 + {\cal O}\left(\frac{\alpha^4}{n^3}\right)\right]
\end{equation}
and therefore the distances between these levels and the highest level in the ``band'' are
\begin{equation}
\Delta M_\alpha = - \frac{15 m}{8\pi\left(n+\frac{3}{4}\right)^3} \alpha^2 + {\cal O}\left(\frac{\alpha^4}{n^3}\right)
\end{equation}
for $\alpha = 1,2,3...$. Figure $3$ gives graphical representation of the ``band'' structure.

A similar analysis can be done for states with other $J^{PC}$ quantum numbers. For brevity, we present here
only the asymptotic expressions for $J^{--}$ state masses:
\begin{itemize}
\item[]
\begin{equation*}
M_{0^{--*^n}} = m\frac{\pi}{2}\left(n+\frac{7}{4} + \frac{\gamma_{2,1} + \gamma_{2,1}}{\pi}\right) + {\cal O}\left(\frac{1}{n}\right)
\end{equation*}
\item[]
\begin{equation*}
M_{2^{--*^n}} = m\frac{\pi}{2}\left(n+\frac{11}{4} + \frac{\gamma_{2,1} + \gamma_{2,2}}{\pi}\right) + {\cal O}\left(\frac{1}{n}\right)
\end{equation*}
\item[]
\begin{equation*}
\ldots\ldots\ldots\ldots\ldots\ldots\ldots\ldots
\end{equation*}
\item[]
\begin{equation}\label{J--asymptotic}
M_{J^{--*^n}} = m\frac{\pi}{2}\left(n+\frac{2J+7}{4} + \frac{\gamma_{2,1} + \gamma_{2,J/2+1}}{\pi}\right) + {\cal O}\left(\frac{1}{n}\right).
\end{equation}
\end{itemize}
and for large values of spin $J$ equation (\ref{J--asymptotic}) further simplifies to
\begin{equation}
M_{J^{--*^n}} = m\frac{\pi}{2}\left(n+J + \frac{7}{2}+ \frac{\gamma_{2,1}}{\pi}\right) + {\cal O}\left(\frac{1}{n}, \frac{1}{J}\right).
\end{equation}
It is interesting to note that asymptotic expressions (\ref{asymptotic}) and (\ref{J--asymptotic}) are reminiscent of empirical observations made in
\cite{OZ}.

Also, if we include all of the spin states, the general structure appears to be reminiscent of the spectrum of a string theory. Although degeneracies are not exact, the bands that we have discussed here appear to be identifiable with the levels of a string spectrum, and there is essentially an exponentially rising density of states. This is the sense in which this theory gives a Hagedorn density of states. The fact that degeneracies are not exact, particularly at the lowest mass levels, indicates that this is not a {\it free} string theory.

\section{Conclusion and Open Questions}


Based on the formalism of ``corner variables'' \cite{knair, bars}
we have presented an evaluation of a quasi-Gaussian wave
functional for Yang-Mills theory in $2+1$ dimensions
containing a very non-trivial kernel compatible with an infinite
number of relativistic point-like resonant states. The masses of
these states are all given as combinations of Bessel function
zeros. The actual numerical values thus obtained are in remarkable
agreement with lattice simulations \cite{teper}. The resulting
picture of glueballs is based on gauge invariant constituent
states, and is rather of an open-string than closed-string (Wilson
loop) like nature. There are a number of outstanding technical issues, and we have been careful to point out throughout the paper the status of each step. 

The perceived success is based heavily on comparisons to the existing lattice data, the most reliable of which are for the lowest lying low spin states. As one goes to higher spin states, there is certainly a lot of physical phenomena that could cloud the picture, such as the mixing of states. Although we have presented some results for non-zero spin, we should perhaps remain cautious, as there are little data to compare to. It is also possible that for some observables, the quasi-Gaussian form of the vacuum wavefunctional becomes insufficient in some way. It would be useful, for example, to construct the actual glueball wavefunctionals and compare their energies to the masses that we have derived here.

In conclusion, we believe that our results demonstrate the importance of
``corner variables'' \cite{knair, bars} in the pure Yang-Mills sector.
As mentioned in the introduction to this paper, these variables
are not only useful in $2+1$ but also in $3+1$ dimensions \cite{withlaurent}.
Many other purely theoretical questions remain: understanding of non-perturbative
renormalization, better understanding of the constituent picture,
exploration of the lattice formulation in terms of ``corner variables'',
understanding of a manifestly covariant formulation {\it etc}.
Obviously on a more pragmatic level, it is important to understand
the inclusion of quarks (and the emergence of mesons and baryons) in this approach in both $2+1$ and $3+1$ dimensions.

\vskip 1cm

{\Large \bf Acknowledgements}

\vskip .2cm

We would like to thank L.~Freidel for many discussions and an illuminating
collaboration. We would also
like to acknowledge conversations with
V.~P.~Nair, S.~Nowling, D. Gross and T.~Takeuchi, and e-mail communications
from E.~Witten, J.~de Boer, I.~Bars and especially M.~Teper.
We also thank numerous colleagues for raising many pertinent questions
during various seminar presentations of the research contained in the present paper.
{RGL} was supported
in part by the U.S. Department of Energy under contract
DE-FG02-91ER40709 and the UIUC Campus Research Board. {DM} and
{AY} were supported in part by the U.S. Department of Energy
under contract DE-FG05-92ER40677 and internal funds provided by
Virginia~Tech. Finally, we thank Perimeter Institute for hospitality.

\section{Appendix A: Regulated Computations}

In this Appendix we want to discuss the eigenstates of the kinetic operator
\begin{equation}
T = m \left(\int_x J^a(x) \frac{\delta}{\delta J^a(x)} + \int_{z,w}\Omega^{ab}(z,w)
\frac{\delta}{\delta J^a(z)} \frac{\delta}{\delta J^b(w)}\right).
\end{equation}
In particular we are interested in the action of $T$ on a special
class of local operators ${\cal O}_n = \int \bar\partial J (\Delta)^n
\bar\partial J$. ${\cal O}_0 = \int \bar\partial J^a \bar\partial
J^a$ has been considered previously in detail in \cite{knair}
and we wish to 
generalize to arbitrary $n$. Since we will be unable to do this infinite number of calculations, we will focus on the general structure of $T{\cal O}_n$. Ultimately, we will find a number of issues in these calculations that we are unable to resolve. The issues include the use of the holomorphic invariant point-splitting regulator and precisely how to remove the regulator, as well as renormalization and normal-ordering issues.\footnote{Another route to this computation is the one followed by
Karabali, Kim and Nair in their computation of the action of the
kinetic operator on $\CO_0$.
Their computation is based on an explicit insertion of Wilson lines
in the point-split form of our $\CO_0$ operator. This may be taken to mean a different basis of operators.
There is apparently a mismatch between our computation and theirs,
which can be traced to the ultralocal form of our $\CO_n$ operators.
The Karabali-Kim-Nair type of computation has not been generalized
in this paper. Obviously it is important to pursue that approach to
check the results presented in this paper.
} The reader should interpret this Appendix as a survey of the issues involved, as well as motivations for eq. (\ref{eq:TOn}); this equation is not derived.

We start with a very important technical point.
The action of the operator
\begin{equation}
T_1 = m \int_x J^a(x) \frac{\delta}{\delta J^a(x)}
\end{equation}
on any local operator-valued function of $J$
is always calculationally straightforward. $T_1$ simply acts as a number operator counting $J$'s.
For example, the action of $T_1$ on ${\cal O}_0 = \int \bar\partial J^a \bar\partial J^a$ is trivial
and we simply get
\begin{equation}
T_1 {\cal O}_0 = 2m {\cal O}_0.
\end{equation}
More generally though, the action of $T_1$ on ${\cal O}_n$ will not result in a holomorphic invariant.

On the other hand the action of the other part of the kinetic operator involving
\begin{equation}
T_2 = m \int_{z,w}\Omega^{ab}(z,w)
\frac{\delta}{\delta J^a(z)} \frac{\delta}{\delta J^b(w)}
\end{equation}
is very subtle. For ${\cal O}_0$ we have
\begin{equation}
\frac{\delta^2 \CO_0}{\delta J^a(z) \delta J^b(w)} = -2\,\delta^{ab}\, {\bar\partial}_z^2 \delta (z-w)
\end{equation}
and so
\begin{equation}\label{eq:T2action}
T_2 \CO_0 = -2m \int_w \delta^{ab}\left[ {\bar\partial}_z^2 \Omega^{ab}(z,w)\right]_{z=w}.
\end{equation}
To compute the right-hand side of Eq.(\ref{eq:T2action}), we use the point-split expression for $\Omega^{ab}(z,w)$
\cite{knair}
\begin{equation}\label{eq:OmegaRegulated}
\Omega^{ab}(z,w) = D_w^{br} \Lambda_{ra}(w,z)
\end{equation}
where $D_w^{br}$ is the adjoint representation of the holomorphic-covariant derivative, and \cite{knair}
\begin{equation}\label{eq:LambdaRegulated}
\Lambda_{ra}(w,z) = \frac{1}{\pi (z-w)} \left[ \delta_{ra} - {\left(H(w) H^{-1}(z, \bar{w})\right)}_{ra}
e^{-\frac{\alpha}{2}} \right] + ...
\end{equation}
with $ \alpha \equiv \frac{|z-w|^2}{\epsilon}$, and $\epsilon\to 0$. Now one easily obtains (summation over repeated index is understood)
\begin{equation}\label{eq:D2zOmega}
\left[ {\bar\partial}_z^2 \Omega^{aa}(z,w)\right]_{z=w} = \frac{c_A}{\pi} \frac{(dim\, G)}{4\pi\epsilon^2}
\end{equation}
and finally
\begin{equation}
T_2\, \CO_0 = - m \int \frac{c_A}{\pi}\frac{(dim\, G)}{2\pi \epsilon^2}.
\end{equation}
From this we see that the only effect of $T_2$ acting on $\CO_0$ is to give a divergent normal-ordering correction. We  may now define
\begin{equation}
:\!\CO_0\!:\ = \int \left[\bar\partial J^a \bar\partial J^a -\frac{c_A}{\pi} \frac{(dim\, G)}{4\pi \epsilon^2}\right]
\end{equation}
and for normal-ordered operator  $:\!\CO_0\!:$ we get
\begin{equation}
T\, :\!\CO_0\!:\ = 2 m :\!\CO_0\!:.
\end{equation}
In what follows we will not keep track of such divergent terms explicitly, however it should be understood that the rest of $\CO_n$ operators
have to be normal-ordered in a similar way.

Let us proceed now to the operator
\beq
\CO_1 =\int  \bar\pa J^a (\Delta)^{ab}\bar\pa J^b =\int \bar\pa J^a D^{ab}\,\bar\pa^2\! J^b.
\eeq
The action of $T_1$ on this is elementary
\begin{equation}
T_1\,\CO_1 = 2m \int \bar\pa J^a (\partial)\, \bar\pa^2\! J^a + 3m \int \bar\pa J^a \left(-i\frac{\pi}{c_A} f^{adb} J^d \right)\bar\pa^2 J^b.
\end{equation}

Note however that this result is not invariant under holomorphic
transformations. What we want to show now is that
when $T_2$ acts on the $3J$-part of $\CO_1$ (the action on the $2J$-part is
trivial and gives only a normal-ordering constant as was discussed
above) it generates an extra term with two $J$'s which restores
holomorphic invariance, {\it i.e.}
\begin{equation}\label{eq:T2}
T_2\, \CO_1 = m \int \bar\pa J^a (\partial)\, \bar\pa^2\! J^a
\end{equation}
and therefore
\begin{equation}
T\, \CO_1 = (T_1 + T_2) \, \CO_1 = 3m \, \CO_1 .
\end{equation}

But before we proceed to show that (\ref{eq:T2}) is  correct we want to make a technical remark which will significantly simplify calculations.
As can be seen from expressions (\ref{eq:OmegaRegulated}) and (\ref{eq:LambdaRegulated}) every time $\bar\partial_z$ acts on $\Omega^{ab}(z,w)$ it actually acts on the exponent  in $\Lambda_{ra}$
and thus pulls out a factor $1/\epsilon$. Therefore expressions like  $\left[ {\bar\partial}_z  \Omega^{ab}(z,w)\right]_{z=w}$,
$\left[ {\bar\partial}_z \partial_w \Omega^{ab}(z,w)\right]_{z=w}$, {\it etc} will lead to divergent normal-ordering terms only ({\it cf}. Eq.(\ref{eq:D2zOmega}), for example).
Such expressions will appear when $\delta/\delta J^a(z)$ partial derivative acts on $J$'s with leading antiholomorphic derivatives, like $\bar\partial J$ or
$\bar\partial^2\! J$ in the case of $\CO_1$ operator. Therefore, as long as we are not interested in explicit form of the normal-ordering terms, we
can consider action of $\delta/\delta J^a(z)$ on ``bare'' $J$'s only. Therefore
\beq
\frac{\delta\CO_1}{\delta J^a(z)}= \left(-i\frac{\pi}{c_A} f^{dae}\right)\, \bar\partial J^d_z\, {\bar\partial}^2\! J^e_z
\eeq
and
\begin{equation}
\frac{\delta\CO_1}{\delta J^a(z)\delta J^b(w)} = \left(-i \frac{\pi}{c_A} f^{bae}\right)\left( {\bar\partial}^2\! J^e_z\:
\bar\partial_z \delta (z-w) - {\bar\partial} J^e_z\: {\bar\partial}_z^2 \delta (z-w)\right)
\end{equation}
from which we immediately obtain
\begin{equation}
T_2\, {\CO_1} = m \int \left(-i \frac{\pi}{c_A} f^{bae}\right) \left({\bar\partial}^2\! J^e \left[{\bar\partial}_w \Omega^{ab}(z,w)\right]_{z=w} -\
{\bar\partial} J^e \left[{\bar\partial}_w^2 \Omega^{ab}(z,w)\right]_{z=w}
 \right).
\end{equation}

To proceed further we need to evaluate $\left[{\bar\partial}_w \Omega^{ab}(z,w)\right]_{z=w}$ and $\left[{\bar\partial}_w^2 \Omega^{ab}(z,w)\right|_{z=w}$.
A straightforward computation gives (we keep only finite terms which do not depend on regularization parameter $\epsilon$)
\begin{eqnarray}
\left[{\bar\partial}_w \Omega^{ab}(z,w)\right]_{z=w}
=
\frac{1}{2\pi}\, \bar\partial\!\left(D J\right)^{ba}
=
\frac{1}{2\pi}\, \partial\bar\partial J^{ba} -
\frac{1}{2\pi}\frac{\pi}{c_A}\,
\bar\partial\!\left(J
J\right)^{ba}\\
\left[{\bar\partial}_w^2
\Omega^{ab}(z,w)\right]_{z=w}
= \frac{1}{2\pi}\, {\bar\partial}^2\!\!\left(D
J\right)^{ba} = \frac{1}{2\pi}\,
\partial{\bar\partial}^2\! J^{ba} -
\frac{1}{2\pi}\frac{\pi}{c_A}\,
{\bar\partial}^2\!\!\left(J
J\right)^{ba}\label{eq:BarD2Omega}
\end{eqnarray}
and
\begin{equation}\label{eq:T2O1}
T_2\, {\CO_1} = m \int \left(-i \frac{1}{2 c_A} f^{bae}\right) \left({\bar\partial}^2\! J^e\: \partial\bar\partial J^{ba} -\
{\bar\partial} J^e\:  \partial{\bar\partial}^2\! J^{ba}
 \right).
\end{equation}
Finally,  we may use the identity $ f^{bae} J^{ba}  = i c_A\, J^e$  to verify that (\ref{eq:T2O1}) is indeed equivalent to (\ref{eq:T2}).

To summarize, what we have shown is that the action of kinetic energy operator  on $\CO_1$ gives a gauge and holomorphic invariant result
\begin{equation}
T\, :\CO_1:  = 3m \, :\CO_1: .
\end{equation}

Finally, let us consider one further example
\begin{equation}
{\cal O}_2 = \int \bar\partial J^a (\Delta^2)^{ab}\, \bar\partial J^b.
\end{equation}
It is convenient to rewrite this as
\begin{equation}
{\cal O}_2 = \int (D\,{\bar\partial}^2\! J)^d(D\,{\bar\partial}^2\! J)^d.
\end{equation}
Proceeding now the same way as we did for ${\cal O}_1$, we obtain
\begin{equation}
\frac{\delta\CO_2}{\delta J^a(z)} = 2\,\frac{\pi}{c_A}\left(-i f^{dac}\, {\bar\partial}^2\! J^c_z\right)\, \left(D\,{\bar\partial}^2\! J\right)^d_z
\end{equation}
and
\begin{eqnarray}
\frac{\delta\CO_2}{\delta J^a(z)\delta J^b(w)} = 2\,\frac{\pi}{c_A}\:\biggl\{ \left(-i f^{dab}\: {\bar\partial}^2_z \delta(z-w)\right) \left(D\,{\bar\partial}^2\! J\right)^d_z +\qquad\qquad\ \\\nonumber
\left(-i f^{dac}\: {\bar\partial}^2\! J^c_z\right)  \left(D^{db}_z\,{\bar\partial}^2_z \delta(z-w)\right)  +\qquad\quad\ \,\\\nonumber
\left(-i f^{dac}\: {\bar\partial}^2\! J^c_z\right) \frac{\pi}{c_A}\,\delta(z-w)\left(-i f^{dbe}\:  {\bar\partial}^2\! J^e_z\right) \biggr\}.
\end{eqnarray}
Now we may write for the action of $T_2$ on $\CO_2$
\begin{eqnarray}\label{eq:T2O2}
T_2\CO_2  = 2 m \frac{\pi}{c_A}\:\biggl\{ \int\left(-i f^{dab} \left[{\bar\partial}^2_w \Omega^{ab}(z,w)\right]_{z=w}\right) \left(D\,{\bar\partial}^2\! J\right)^d +\qquad\quad\ \ \ \ \!\\\nonumber
\int \left(-i f^{dac}\: {\bar\partial}^2\! J^c\right)  \left(\left[D^{db}_w\,{\bar\partial}^2_w \Omega^{ab}(z,w)\right]_{z=w}\right)  +\qquad\quad\ \\\nonumber
\int \left(-i f^{dac}\: {\bar\partial}^2\! J^c\right) \frac{\pi}{c_A}\left[\Omega^{ab}(z,w)\right]_{z=w} \left(-i f^{dbe}\:  {\bar\partial}^2\! J^e\right) \biggr\}.
\end{eqnarray}

Since $\left[{\bar\partial}^2_w \Omega^{ab}(z,w)\right]_{z=w}$ was computed previously in (\ref{eq:BarD2Omega}), we immediately obtain for the first term in (\ref{eq:T2O2})
\begin{equation}\label{eq:O2firstterm}
2 m \,\frac{\pi}{c_A} \int\left(-i f^{dab} \left[{\bar\partial}^2_w \Omega^{ab}(z,w)\right]_{z=w}\right) \left(D\,{\bar\partial}^2\! J\right)^d_z\ =\ 
m \int \left(\partial{\bar\partial}^2\! J\right)^d \left(D\,{\bar\partial}^2\! J\right)^d.
\end{equation}
Similarly for the last term in (\ref{eq:T2O2}) we need expression for $\left[\Omega^{ab}(z,w)\right]_{z=w}$ which is
\begin{equation}
\left[\Omega^{ab}(z,w)\right]_{z=w} = \frac{1}{2\pi} \left(D J\right)^{ba} = \frac{1}{2\pi} \partial J^{ba} -
\frac{1}{2\pi}\frac{\pi}{c_A}\left(J J\right)^{ba}
\end{equation}
and therefore
\begin{equation}\label{eq:O2thirdterm}
2 m \left(\frac{\pi}{c_A}\right)^2\!\! \int \left(-i f^{dac}\: {\bar\partial}^2\! J^c_z\right)\left[\Omega^{ab}(z,w)\right]_{z=w} \left(-i f^{dbe}\:  {\bar\partial}^2\! J^e_z\right)
= -m \frac{1}{\pi}\left(\frac{\pi}{c_A}\right)^2\!\! \int \left( {\bar\partial}^2\! J{\bar\partial}^2\! J D J \right)^{aa}
\end{equation}
where summation over the repeated index $a$ on the right-hand side of this expression  is implied (this essentially amounts to taking trace in adjoint
representation).
Finally, for the second term in (\ref{eq:T2O2}) we will need
\begin{equation}
\left[D^{db}_w\,{\bar\partial}^2_w \Omega^{ab}(z,w)\right]_{z=w} = \frac{1}{2\pi}\left\{D{\bar\partial}^2 D J - \frac{1}{3}{\bar\partial}^2 D^2 J\right\}^{da}
\end{equation}
from which we obtain
\begin{equation}\label{eq:O2secondterm}
2 m \frac{\pi}{c_A}\int \left(-i f^{dac}\: {\bar\partial}^2\! J^c_z\right)  \left(\left[D^{db}_w\,{\bar\partial}^2_w \Omega^{ab}(z,w)\right]_{z=w}\right)
= - \frac{m}{c_A}\int \left({\bar\partial}^2 J  \left\{D{\bar\partial}^2 D J - \frac{1}{3}{\bar\partial}^2 D^2 J\right\}\right)^{aa}.
\end{equation}
Now if we collect all the results in equations
(\ref{eq:O2firstterm}), (\ref{eq:O2thirdterm}) and
(\ref{eq:O2secondterm}) together and after some simple but
somewhat lengthy transformations  we obtain the final expression
for the action of $T_2$ operator on $\CO_2$ state
\begin{eqnarray}
T_2\, \CO_2 = \frac{m}{3}\, \biggl\{ 5 \int \left(\partial{\bar\partial}^2\! J\right)^a \left(\partial{\bar\partial}^2\! J\right)^a - \qquad\qquad\quad\ \\\nonumber
4\int \left(\partial{\bar\partial}^2\! J\right)^a \left(\frac{\pi}{c_A}\, [J, {\bar\partial}^2\! J]\right)^a - \:\,\qquad \\\nonumber
\int \left(\frac{\pi}{c_A}\, [J, {\bar\partial}^2\! J]\right)^a \left(\frac{\pi}{c_A}\, [J, {\bar\partial}^2\! J]\right)^a - \!\!\!\\\nonumber
\int \frac{\pi}{c_A} \left([\bar\partial J, {\bar\partial}^2 J]\right)^d \left(D \bar\partial J\right)^d \biggr\}\quad\qquad
\end{eqnarray}
which combined with the straightforward result for $T_1 \CO_2$
\begin{eqnarray}
T_1\, \CO_2 = m\, \biggl\{2 \int \left(\partial{\bar\partial}^2\! J\right)^a \left(\partial{\bar\partial}^2\! J\right)^a - \qquad\qquad\quad\ \,\\\nonumber
6 \int \left(\partial{\bar\partial}^2\! J\right)^a \left(\frac{\pi}{c_A}\, [J, {\bar\partial}^2\! J]\right)^a +\qquad\ \:\\\nonumber
4 \int \left(\frac{\pi}{c_A}\, [J, {\bar\partial}^2\! J]\right)^a \left(\frac{\pi}{c_A}\, [J, {\bar\partial}^2\! J]\right)^a\biggr\}
\end{eqnarray}
finally gives
\begin{equation}\label{eq:TO2final}
T\, \CO_2  = \left(4-\frac{1}{3}\right)m \CO_2 + \frac{m}{3}\frac{\pi}{c_A} \int \left([\bar\partial J, {\bar\partial}^2 J]\right)^d \left(D \bar\partial J\right)^d.
\end{equation}
As a consistency check one may notice that we obtain a gauge and holomorphically invariant expression as expected.

A few comments about this result are in order, as it is not of the form that we have expected. First of all we see that 
an extra gauge invariant operator $\int \left([\bar\partial J,
{\bar\partial}^2 J]\right)^d \left(D \bar\partial J\right)^d$
mixes with $\CO_2$. Such operator mixing is certainly expected to
be the general feature of the theory and therefore we may write
\begin{equation}\label{eq:TOnG}
T \CO_n  = E_{n} {\CO}_n +\ldots
\end{equation}
where ``$\ldots$'' indicate operators which mix with $\CO_n$ under
the action of $T$. It should be noted, however, that such
operators, even though they have the same mass dimension as
$\CO_n$, are of higher order in $\bar\pa J$. By this we mean that only
$\CO_n$ in equation (\ref{eq:TOnG}) contains terms quadratic in
$\bar\pa J$ while the rest of the operators have at least three $\bar\pa J$'s. This means, in
particular, that such operator mixing has no influence on
derivation of Schr\"odinger equation presented in this paper and
therefore detailed knowledge of ``$\ldots$'' terms in equation
(\ref{eq:TOnG}) is not really necessary for our purposes.

More worrisome is the numerical coefficient of ${\CO}_2$ in (\ref{eq:TO2final}).
As can be seen from
(\ref{eq:TO2final}), we may think of $\CO_2$ as an eigenstate (up
to an extra operator that mixes in) of kinetic energy with
eigenvalue $(4-1/3)m$. 

We believe, however, that this result is not complete and the correct spectrum for the kinetic energy should be equidistant
\begin{equation}\label{eq:expected}
T \CO_n  = (n+2) m \CO_n +\ldots.
\end{equation}
The reason why we have not obtained the expected answer in the case of ${\cal O}_2$ (and presumably for ${\cal O}_{n>2}$) may be that we have not taken proper account of the nonlocal character of the theory. Thus the calculation that we have outlined above needs to be reconsidered carefully. We do not at this time have a consistent understanding of all of these issues as the required calculations are tedious, but we present remarks below. In the text of the paper, we use eq. (\ref{eq:expected}) without further apologies.

\subsection{Further discussion: regulator/renormalization issues}

In deriving results for $\CO_0$, $\CO_1$ and $\CO_2$ operators we
have been using expression for $\Omega^{ab}(z,w)$ with
$\Lambda_{ra}(w,z)$ given in equation (\ref{eq:LambdaRegulated}).
This expression for $\Lambda_{ra}(w,z)$ is in fact a simplified
form of the exact regulator \cite{knair}
\begin{equation}\label{eq:LambdaRegulatedExact}
\Lambda_{ra}(w,z) = \frac{1}{\pi (z-w)} \left[ \delta_{ra} - {\left(H(w) H^{-1}(u, \bar{w})H(u)H^{-1}(z, \bar{u})\right)}_{ra}
e^{-\frac{\alpha}{2}} \right] + ...
\end{equation}
where $u=\frac{1}{2}(z+w)$ and
$\bar{u}=\frac{1}{2}(\bar{z}+\bar{w})$. It is easy to see that
(\ref{eq:LambdaRegulated}) can be obtained from this expression if
we put $\bar{z}=\bar{w}$ in ${\left(H(w) H^{-1}(u,
\bar{w})H(u)H^{-1}(z, \bar{u})\right)}_{ra}$ . Therefore
(\ref{eq:LambdaRegulated}) is equivalent to
(\ref{eq:LambdaRegulatedExact}) up to the $O(\bar{z}-\bar{w})$
terms. Part of the problem uncovered in the previous section then may have to do with subtleties 
involving the assumed form of $\Lambda_{ra}$. It seems that for the study of the action of $T_2$ on
local operators, such simplification is not justified and exact
expression  (\ref{eq:LambdaRegulatedExact}) should be used.
Although it becomes extremely difficult to do calculations with
this $\Lambda_{ra}(w,z)$, we have considered this effect on $\CO_0$ and $\CO_1$. There appear to be additional contributions (of the form $\epsilon/\epsilon$) even to their eigenvalues
\begin{eqnarray}
T\,\CO_0 \sim 2\left(1-\frac{1}{16}\right)m \CO_0\\\nonumber
T\,\CO_1 \sim 3\left(1-\frac{3}{32}\right)m \CO_1.
\end{eqnarray}
What does this effect illustrate? There are certainly very subtle effects coming from the point-splitting regulator. We have not been able to arrive at a consistent method for dealing with these effects. In addition though, we should note that these are all formal expressions -- we have not carefully defined the operators or the coupling, particularly in the infrared. It is not clear to us that such a procedure is known in this or any other (non-trivial) theory.

Let us be a little more explicit about the issue raised here. In performing computations
for $\CO_0$, $\CO_1$ and $\CO_2$ we have kept track of $\epsilon$-independent terms only.
In general, however, we may write
\begin{equation}\label{eq:Tepsilon}
\begin{array}{c}
T\, \CO_0 = \ldots + T_{00}\,\CO_0 + T_{10}\, \epsilon\,\CO_1 + T_{20}\, \epsilon^2\,\CO_2 +\ldots\\
T\, \CO_1 = \ldots + T_{11}\,\CO_1 + T_{21}\, \epsilon\,\CO_2 + T_{31}\, \epsilon^2\,\CO_3 +\ldots\\
T\, \CO_2 = \ldots + T_{22}\,\CO_2 + T_{32}\, \epsilon\,\CO_3 + T_{42}\, \epsilon^2\,\CO_4 +\ldots\\
\ldots\ldots\ldots\ldots\ldots\ldots\ldots\ldots\ldots
\end{array}
\end{equation}
or in matrix notation
\begin{equation}
T\, \CO_i = T_{j i}\, \epsilon^{j-i}\, \CO_j.
\end{equation}
This is a schematic expression  based solely on dimensional
analysis (remember that $[\CO_n] = {\rm (mass)}^{4+2n}$). It does
not include, for example, the effects of mixing with operators
other than $\CO_n$. $T_{ij}$ here is an infinite dimensional
number matrix which can be thought of as a representation of
kinetic energy operator in the basis provided by operators
$\CO_n$. Evaluating the spectrum of $T$ would then certainly amount
to diagonalizing $T_{ij}$. In some sense what we have done above
is have found diagonal elements $T_{nn}$ of this matrix and
associated them with the spectrum of kinetic operator. In other
words if one takes naive $\epsilon\to 0$ limit in
(\ref{eq:Tepsilon}) it looks like all terms with positive powers
of $\epsilon$ go to zero, $T_{ij}$ matrix becomes triangular and,
consequently, diagonal elements give the spectrum. The proper
procedure, however, would be to diagonalize $T_{ij}$ first and
take $\epsilon \to 0$ limit afterwards. Formally we may write the
$n$-th eigenstate as a linear combination of an infinite number of
local operators $\CO_n$
\begin{equation}\label{eq:Qdefined}
Q^{(n)} = c^{(n)}_i\,\epsilon^{i-n}\,\CO_i
\end{equation}
with some unknown numeric coefficients $c^{(n)}_i$. The eigenvalue equation
\begin{equation}
T\, {\cal Q}^{(n)} = E^{(n)} {\cal Q}^{(n)}
\end{equation}
becomes now the infinite-dimensional matrix equation
\begin{equation}
T_{j i}\, c^{(n)}_i = E^{(n)}\,  c^{(n)}_j
\end{equation}
from which in principle we could find the spectrum. Of course, to do this practically would be a very difficult task.
Note, however, that the $\epsilon$-dependence completely drops out from this expression. This means, in particular, that the spectrum of $T$ does not depend on the regularization
parameter $\epsilon$ and we may take the formal limit $\epsilon \to 0$ in (\ref{eq:Qdefined}) after diagonalization. At any finite value of $\epsilon$ the true eigenstates
${\cal Q}^{(n)}$ of $T$ are certainly nonlocal, however we expect that they will collapse into local states as $\epsilon \to 0$, {\it i.e.}
\begin{equation}
{\cal Q}^{(n)} \stackrel{\epsilon\to\, 0}\longrightarrow\ :\!\CO_n\!:.
\end{equation}
Therefore for any practical purposes, like writing Schr\"odinger equation for the full Hamiltonian ${\cal H}$, we may think of the eigenstates of
kinetic energy in terms of local operators $:\!\CO_n\!:$ and we may now really write
\begin{equation}
T\, :\!\CO_n\!: \stackrel{\epsilon \to 0} = E^{(n)} :\!\CO_n\!: +\ldots
\end{equation}
It should be noted however that this expression, even though it looks completely local, knows in fact about nonlocal character of the theory. This nonlocal
information is contained in the spectrum $E^{(n)}$ which should be obtained by solving eigenvalue equations with nonlocal ansatz (\ref{eq:Qdefined}).

To conclude this discussion, we would like to consider a
simple example which further illustrates the intuitive arguments
presented above. In \cite{knair} the following nonlocal
generalization of $\CO_0$ was proposed
\begin{equation}
{\cal Q}^{(0)}_\epsilon = \int_{x,y} \frac{e^{-\frac{|x-y|^2}{\epsilon}}}{\pi \epsilon}\ \bar\partial J^a(x) \left[H(x,\bar{y})H^{-1}(y)\right]^{ab} \bar\partial J^a(x).
\end{equation}
This state is essentially made of two $\bar\partial J$'s connected by a straight Wilson line. It is not difficult now to expand this in terms
of local operators with the result given by
\begin{equation}
{\cal Q}^{(0)}_\epsilon = {\cal R}_0 + \epsilon {\cal R}_1 + \frac{1}{2!}\epsilon^2 {\cal R}_2 + \frac{1}{3!}\epsilon^3 {\cal R}_3 +\ldots
\end{equation}
where
\begin{equation}
{\cal R}_n = \int \bar\partial J^a \left(D^n {\bar\partial}^n\right)^{ab} \bar\partial J^b.
\end{equation}
These ${\cal R}_n$ operators are not quite the same as $\CO_n$
(except ${\cal R}_0\equiv\CO_0$ and ${\cal R}_1\equiv\CO_1$),
however they coincide with the latter to quadratic order in $J$.
Therefore from this expansion we see that ${\cal
Q}^{(0)}_\epsilon$ is of the same type as we need to diagonalize
$T$ according to (\ref{eq:Qdefined}). Of course, this simple
operator is not an eigenstate of $T$, however we may see the other essential ingredient of
the above analysis, namely the fact that in the continuum
$\epsilon\to 0$ limit, ${\cal Q}^{(0)}_\epsilon$ collapses into
local operator $\CO_0$.

\section{Appendix B: Vacuum Energy}

Here we discuss the vacuum energy. For this purpose, we need to evaluate the Schr\"odinger equation in the form
\beq
{\cal H}_{KN}\Psi_0= (E_0+\ldots)\Psi_0
\eeq
where $E_0$ consists of terms that survive as $J^a\to 0$. Since the Hamiltonian involves at most two derivatives with respect to $J$, we can drop any $J$-dependence of the kernel for this computation and work with
\beq
\Psi_0\simeq {\rm exp}\left({-\frac{\pi}{2 c_A m^2}\int\bar\partial J^a K[\partial\bar\partial/m^2] \bar\partial J^a}\right).
\eeq
We then compute
\beq
\frac{\delta}{\delta J^a(z)}\Psi_0 = \frac{\pi}{c_A m^2}\left(\bar\partial\, K\!\left[\partial\bar\partial/m^2\right] \bar\partial J\right)^a_z \Psi_0
\eeq
and
\beq
\frac{\delta^2}{\delta J^b(w)\delta J^a(z)}\Psi_0 = \frac{\pi}{c_A m^2}\, \left(K\!\left[\partial\bar\partial/m^2\right]\! \bar\partial^2\right)_z \delta^{(2)}(z-w)\delta^{ab} \Psi_0 +\ldots
\eeq
and so
\begin{equation}
m\int_{w,z}\Omega^{ab}(z,w)\frac{\delta^2}{\delta J^b(w)\delta J^a(z)}\Psi_0=
\left(\int_z {\cal E}_0+\ldots\right)\Psi_0.
\end{equation}
We then deduce that the vacuum energy density is
\begin{equation}
{\cal E}_0=\frac{\pi}{c_A m}\,\delta_{ab}\left. \left(K\!\left[\partial\bar\partial/m^2\right] {\bar\partial}^2\right)_z \Omega^{ab}(z,w)\right|_{w=z}.
\end{equation}
To proceed, we write out the regulated $\Omega^{ab}$ keeping only those terms that are present in the $J^a\to 0$ limit. This is particularly simple, and we find
\begin{equation}
\Omega^{ab}_{reg}(z,w)=\frac{c_A}{\pi}(D_w\Lambda(w,z))^{ba}=\frac{c_A}{\pi}\,\delta^{ab}\,\partial_w \frac{1}{\pi (z-w)} \left(1-e^{-|z-w|^2/2\epsilon}\right).
\end{equation}
In the expression for the vacuum energy, there is a leading $\bar\partial_z^2$; performing one of these derivatives on $\Omega^{ab}_{reg}$, we find
\begin{eqnarray}\nonumber
{\cal E}_0&=&\frac{dim\,G}{m}\left. \left(K\!\left[\partial\bar\partial/m^2\right] {\bar\partial}\right)_z \partial_w \frac{1}{2\pi\epsilon}e^{-|z-w|^2/2\epsilon}\right|_{w=z}\\
&=& - m(dim\,G)\ \left. \left( K\!\left[\frac{\partial\bar\partial}{m^2}\right] \frac{\partial\bar\partial}{m^2} \right)_z\frac{1}{2\pi\epsilon}e^{-|z-w|^2/2\epsilon}\right|_{w=z}.
\end{eqnarray}
This expression is horribly divergent. Let us proceed however by
considering the expansion for $K$ \beq K[x]\,x =\sum_{n=0}^\infty
c_n x^{n+1}. \eeq We then find \beq {\cal E}_0= -\frac{(dim\,G)\
m}{2\pi\epsilon} \sum_{n=0}^\infty c_n (n+1)! \left(
-\frac{1}{2\epsilon m^2}\right)^{n+1}. \eeq Note that the
appearance of divergences to arbitrary order here comes directly
from the fact that the vacuum wave-functional contains operators
of arbitrarily high dimension. This basic fact plagues all such
computations. In the case of the vacuum energy, it is possible to
resum this series and extract sensible results (in particular, the
asymptotic behavior is consistent with the UV perturbation
theory). The basic observation is that
\beq (n+1)!\ t^{-2-n}=\int_0^\infty dx\ x^{n+1} e^{-tx}
\eeq and so we have
\begin{eqnarray}\nonumber
{\cal E}_0&=& \frac{(dim\,G)\ m^3}{\pi} \sum_{n=0}^\infty c_n (-1)^n \left. \int_0^\infty dx\ x^{n+1} e^{-t x} \right|_{t=2\epsilon m^2}\\
&=&\frac{(dim\,G)\ m^3}{\pi} \left. \int_0^\infty dx\ xK(-x) e^{-t x}\right|_{t=2\epsilon m^2}.
\end{eqnarray}
Thus the vacuum energy may be written as the Laplace transform of $xK(-x)$. Removing the regulator corresponds to taken the asymptotic $t=2\epsilon m^2\to 0$ limit.

\subsection{Evaluation of ${\cal E}_0$}

To proceed further, we need to supply information about $K$. We will consider several examples, culminating with the kernel (\ref{eq:normvacuum}).

\subsubsection{$K\equiv 1$}

The simplest choice is to consider $K\equiv 1$, which corresponds to replacing the true $K$ by its IR limit. In this case, the Laplace transform is elementary, and we obtain
\beq
{\cal E}^{K\equiv 1}_0= \frac{dim\,G}{4\pi m\epsilon^2}
\eeq
which is quartically divergent (in powers of a momentum cutoff).

\subsubsection{Massless Boson}

If we consider the wavefunctional corresponding to a free massless boson, we have $K(-x)=1/\sqrt{x}$. Again, the Laplace transform is elementary, and we obtain
\beq
{\cal E}^{particle}_0= \frac{dim\,G}{4\sqrt{2\pi} \epsilon^{3/2}}.
\eeq
This is also the UV limit of the true $K$; we see here that the scale $m$ drops out of this result, as it should.

\subsubsection{Bessel}

Finally, let us consider the kernel (\ref{eq:normvacuum}). $K(-x)$
can be written as \beq
K(-x)=\frac{1}{\sqrt{x}}\frac{I_2(4\sqrt{x})}{I_1(4\sqrt{x})} \eeq
and so we need to compute \beq {\cal E}^{YM}_0=\frac{(dim\,G)\
m^3}{\pi} \int_0^\infty dx\
\sqrt{x}\,\frac{I_2(4\sqrt{x})}{I_1(4\sqrt{x})}\, e^{-t x}. \eeq
Because we are interested in the small $t$ limit, it is sufficient
to consider the asymptotic $x\to\infty$ behavior of the integrand
\beq \sqrt{x}\frac{I_2(4\sqrt{x})}{I_1(4\sqrt{x})}\simeq
\sqrt{x}-\frac{3}{8}+\frac{3}{128}\frac{1}{\sqrt{x}}+ {\cal
O}(1/x). \eeq We thus obtain \beq {\cal E}^{YM}_0=\frac{dim\ G
}{4\sqrt{2\pi}\epsilon^{3/2}}\left[ 1-\frac{3m\sqrt{\epsilon}
}{2\sqrt{2\pi} }+\frac{3m^2\epsilon }{32}\right]+{\cal O}(m^3\log
(\epsilon m)). \eeq Note that the leading divergence is the same
as the UV result, while the subleading terms depend on the
detailed form of the kernel.

\section{Appendix C: Vacuum Solutions}

Assuming the validity of the Riccati equation (\ref{eq:Riccati})
\begin{equation}\label{eq:SchrEqA}
 -K - \frac{L}{2} \frac{d}{dL}[K(L)] + L K^2 +1 = 0
\end{equation}
we now consider its general solution. This is a non-linear differential equation but it is easily solved by
substituting $K=-y'/2y$ to obtain
\begin{equation}
L y'' +2 y' + 4y =0.
\end{equation}
Then, writing $y=f(x)/\sqrt{L}$ with $x=4\sqrt{L}$ leads to the Bessel equation
\begin{equation}
f''(x) +f'(x)/x +(1-1/x^2)f(x)=0
\end{equation}
with general solution
\begin{equation}
y = c_1 \frac{1}{\sqrt{L}} J_1 (4\sqrt{L}) + c_2 \frac{1}{\sqrt{L}} Y_1 (4\sqrt{L}).
\end{equation}
Using the standard recursive formulae (with similar formulae for $Y_\nu$)
\begin{equation}\label{eq:besselrelns}
\begin{array}{ccl}
J_{\nu - 1}(u) + J_{\nu + 1}(u) &=& \frac{2 \nu}{u}J_{\nu}(u) \\
&&\\
J_{\nu - 1}(u) - J_{\nu + 1}(u) &=& 2 J'_{\nu}(u)
\end{array}
\end{equation}
the general solution of (\ref{eq:SchrEqA}) takes the form
\begin{equation}
K(L)=\frac{1}{\sqrt{L}}\frac{CJ_2(4\sqrt{L})+Y_2(4\sqrt{L})}{CJ_1(4\sqrt{L})+Y_1(4\sqrt{L})}.
\end{equation}
Here we have a one-parameter family of solutions, parameterized by $C$. The solution quoted in the text (\ref{eq:normvacuum}) corresponds to $C\to\infty$. To see why we should take this value, we study the asymptotics of the solutions, paying close attention to the normalizability of the solution. It is particularly satisfying that the only normalizable solution corresponds to the confining vacuum.

To study the asympotics, we note that $L=\pa\bar\pa/m^2\to -k^2/4m^2$ is negative (choice of square root doesn't matter), and so we need the identities
\beqn\nonumber
J_1(iz)= iI_1(z),\ \ && \ \ J_2( iz)=-I_2(z),\\
Y_1(iz)=-I_1(z)+\frac{2i}{\pi}K_1(z),\ \ && \ \
Y_2(iz)=-iI_2(z)+\frac{2}{\pi}K_2(z)
\eeqn
giving
\begin{equation}
K(-|L|)=\frac{1}{\sqrt{|L|}}\frac{C'I_2(4\sqrt{|L|})-K_2(4\sqrt{|L|})}{C'I_1(4\sqrt{|L|})+K_1(4\sqrt{|L|})}
\eeq
where $C'=\pi(C+i)/2$. We will consider real $C'$ in what follows (this corresponds to real wavefunctionals), but this is not a crucial simplification: the final conclusion will be unchanged.

Now, asymptotically $|L|\to\infty$, this function behaves
differently, depending on whether $C'=0$ or not, because only for
that value of $C'$ will the $K_n$'s play a role:
\begin{equation}
K(-|L|)\ \substack{|L|\to\infty\\ \longrightarrow} \left\{\begin{array}{rl}+\frac{1}{\sqrt{|L|}},& for\ |C'|\neq 0\cr
-\frac{1}{\sqrt{|L|}},& for\ |C'|= 0.\end{array}\right.
\end{equation}
For small $|L|$ on the other hand, the $K_n$'s will dominate, unless $C'\to\pm\infty$. Thus we find
\begin{equation}
K(-|L|)\ \substack{|L|\to 0\\ \longrightarrow} \left\{\begin{array}{rl}-\frac{1}{2|L|},& for\ C'\neq \pm\infty\cr
1,& for\ C'= \pm\infty.\end{array}\right.
\end{equation}
Thus there are three types of solutions, in terms of their asymptotics. Let us discuss them in turn:

\begin{enumerate}
\item $C'=0$: for this solution $K(-|L|) < 0$ for each value of $|L|$ meaning that this solution is completely non-normalizable (modes of all
momenta are non-normalizable);
\item $C'\neq 0, C'\neq \pm\infty$: for these solutions $K(-|L|)$ changes sign from negative to positive at some finite value of $|L|$ meaning that
it is normalizable in the ultraviolet but non-normalizable in the infrared;
\item $C'=\pm\infty$: for this solution  $K(-|L|)$ is everywhere positive meaning that this solution is completely normalizable.
\end{enumerate}

To summarize, as one can see from this discussion the only solution of Riccati equation (\ref{eq:SchrEqA}) which is normalizable in both UV and IR corresponds
to $C'=\pm\infty$ and is given by
\begin{equation}
K(L) = \frac{1}{\sqrt{L}} \frac{J_2(4 \sqrt{L})}{J_1(4 \sqrt{L})}.
\end{equation}

\section{Appendix D: Bound states}

For a free scalar, the Hamiltonian is
\begin{equation}
{\cal H}_s =  \frac{1}{2}\int \left( - \frac{\delta^2}{\delta \phi^2} +\phi(m^2 - \vec\nabla^2)\phi\right)
\end{equation}
and the vacuum state is
\begin{equation}
\Psi_0 = {\rm exp}\left(-\frac{1}{2}\int \phi(x) \left[\sqrt{k^2+m^2}\right]_{x-y}\phi(y)\right).
\end{equation}
The kernel is therefore $K_s(\vec k) = \sqrt{\vec k^2+m^2}$ and the inverse kernel is
\begin{equation}
K_s^{-1}(\vec k) = \frac{1}{\sqrt{\vec k^2+m^2}}.
\end{equation}
In 2+1 dimensions, we can also deduce, for spatial separation
\begin{equation}\label{eq:prop}
D(\vec x-\vec y)\equiv\langle \phi(\vec x)\phi(\vec y)\rangle=\frac{1}{2\pi |\vec x-\vec y|}\,e^{-m|\vec x-\vec y|}
\end{equation}
which is equivalent to a propagator with single pole. This functional form is what we found in
the factorization of the $0^{++}$ correlator (\ref{spin0correlator}).

In our case, we can also see this physics in momentum space.
Now, in our case the inverse kernel takes the form
\begin{equation}
K^{-1}(\vec k) = \ldots + \frac{1}{{\vec k^2+M_i^2}}+\ldots.
\end{equation}

Compute now the Fourier transform of $(K^{-1}(\vec x-\vec y))^2$. We claimed in the paper that it had the form of (\ref{eq:prop}).
\begin{equation}
(K^{-1})^2[\vec k] = \int \frac{d^2 \vec p}{(2\pi)^2}\ K^{-1}(\vec p)\,K^{-1}(\vec k-\vec p).
\end{equation}
Given the expansion of $K^{-1}(\vec k)$ we may rewrite this as
\begin{equation}
(K^{-1})^2[\vec k] =\ \ldots + \int \frac{d^2 \vec p}{(2\pi)^2}\, \frac{1}{{\vec p}^2+M_i^2}\, \frac{1}{{(\vec k-\vec p)^2+M_j^2}} +\ldots
\end{equation}
so essentially we have to compute a simple two-dimensional scalar loop integral. This is a simple computation and, for example, for equal constituent
masses $i=j$  we get
\begin{equation}
(K^{-1})^2[\vec k]\ \sim\  \ldots + \frac{1}{\sqrt{k^2 + 4 M_i^2}}\frac{{\rm ArcTan} \frac{k/2}{\sqrt{k^2 + 4 M_i^2}}}{ 2\pi k }+\ldots.
\end{equation}
Now we immediately see that this expression describes a physical single-particle state of mass $M= 2M_i$: the leading square root factor is the same as for a free scalar, as above, while the multiplicative function of momentum does not seem to have any unwanted analytic properties.

\end{document}